# OBSERVATORIES IN SPACE

**Catherine Turon**
*GEPI-UMR CNRS 8111, Observatoire de Paris, Section de Meudon, 92195 Meudon, France*

**Contents**

1. Introduction
2. The impact of the Earth atmosphere on astronomical observations
3. High-energy space observatories
4. Optical-Ultraviolet space observatories
5. Infrared, sub-millimeter and millimeter-space observatories
6. Gravitational waves space observatories
7. Conclusion

**Summary**

Space observatories are having major impacts on our knowledge of the Universe, from the Solar neighborhood to the cosmological background, opening many new windows out of reach to ground-based observatories. Celestial objects emit all over the electromagnetic spectrum, and the Earth's atmosphere blocks a large part of them. Moreover, space offers a very stable environment from where the whole sky can be observed with no (or very little) perturbations, providing new observing possibilities. This chapter presents a few striking examples of astrophysics space observatories and of major results spanning from the Solar neighborhood and our Galaxy to external galaxies, quasars and the cosmological background.

## 1. Introduction

Observing the sky, charting the places, motions and luminosities of celestial objects, elaborating complex models to interpret their apparent positions and their variations, and figure out the position of the Earth – later the Solar System or the Galaxy – in the Universe is a long-standing activity of mankind. It has been made for centuries from the ground and in the optical wavelengths, first measuring the positions, motions and brightness of stars, then analyzing their color and spectra to understand their physical nature, then analyzing the light received from other objects: gas, nebulae, quasars, etc. It was not before 1930-1940 that a new window to the Universe was opened with the discovery that some celestial sources emitted not only in visible wavelengths but also in radio wavelengths: the centre of the Galaxy and its spiral arms, the Sun and supernovae remnants.

The story of the word "satellite" is a long one as it was introduced by Kepler in 1610 when he was observing the moons of Jupiter, just discovered by Galileo. Even though the first idea to observe celestial objects from space may originate from Jules Vernes with his book *From the Earth to the Moon*, written in 1865, the first scientific work explaining how to send an "artificial satellite" in orbit around the Earth is probably the book of Konstantin Eduardovich Tsiolkovsky, T*he Exploration of Cosmic Space by Means of Reaction Devices*, published in 1903. Half a century later, the study of the Earth environment during a period of solar maximum activity was the occasion to create the International Geophysical Year (July 1957 to December 1958), and its committee urged the participating countries to use the observing possibilities of artificial satellites to better reach their scientific goal, the study of the upper atmosphere. Taking benefit of the technology developments driven by the need for weapons during World War II, the first launches took place late 1957 and early 1958: Sputnik 1 and 2 launched by the Soviet Union on 4 October and 3 November 1957, and Explorer 1 launched by the United States of America on 31 January 1958. Even though a scientific impulse was at the origin of these launches, they were clearly the result of a political race between the two Superpowers. Nevertheless, the scientists - and especially the astronomers – were rapidly the first the exploit the new possibilities offered by space techniques: as soon as 1961, many planetary probes (Mariner, Ranger, Venera, Luna, etc.) were launched to the Moon, Mars and Venus by both the USA and USSR.

The astronomers also realized very soon the importance of installing instruments on-board satellites with the goal of observing celestial sources emitting in wavelengths of the electromagnetic spectrum unobservable from the ground because they are blocked by the atmosphere. Explorer 11 (shown in Figure 1, left) was the first gamma-ray observatory. Launched by the recently created NASA (the USA National Aeronautics and Space Administration) on 27 April 1961, it observed 22 events attributed to cosmic gamma-rays all over the sky. Uhuru (shown in Figure 1, right), also launched by NASA on 12 December 1970, was the first Earth-orbiting mission entirely dedicated to X-ray astronomy. It observed 339 sources, published by W. Forman et al. in 1978 as "The fourth UHURU catalog of X-ray sources". These sources were mainly identified to binary stellar systems in the Milky Way, supernova remnants, Seyfert galaxies and clusters of galaxies. Besides these pioneering missions especially designed to explore the high-energy domain, many detections of celestial high-energy sources were made by satellites launched for many other reasons: observation of the Sun or watch for countries violating the interdiction of atmospheric testing of nuclear weapons (!).

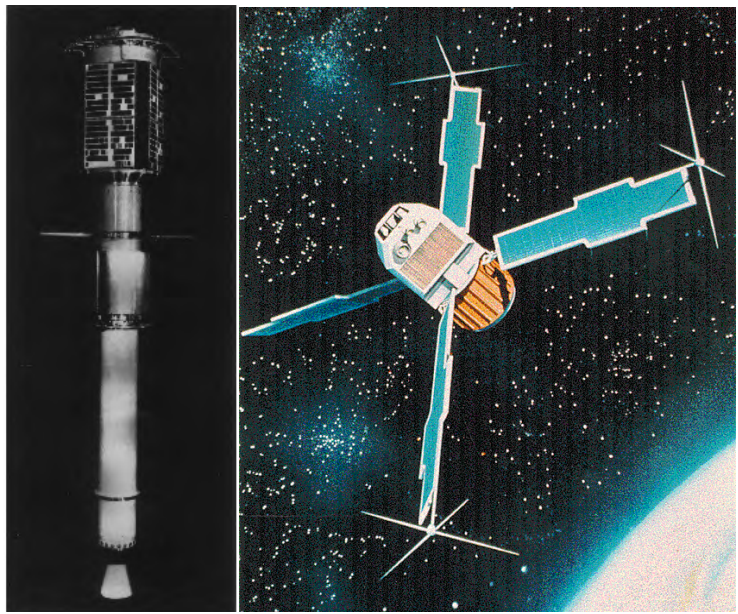

Figure 1. Left: Explorer 11, first gamma-ray satellite, launched by NASA on 27 April 1961 (from http://heasarc.gsfc.nasa.gov/docs/heasarc/missions/explorer11.html). Right: Artist view of Uhuru, first Earth-orbiting satellite entirely dedicated to X-ray astronomy, launched by NASA on 12 December 1970 (from http://heasarc.gsfc.nasa.gov/docs/uhuru/uhuru.html).

After these pioneering high-energy missions, astronomy benefited from space environment to collect observations all over the electromagnetic radiation range. Indeed, space has major advantages for astronomy:

- Space observations are free from the absorption caused by the Earth's atmosphere. Indeed, the Earth's atmosphere is opaque to most of the electromagnetic radiation spectrum, with the exception of the visible light, some infrared (IR) and ultraviolet (UV) wavelengths, and most of the radio domain. Astronomical objects emit in the whole range of the electromagnetic spectrum and going to space opens many new windows to the Universe.
- Space offers a very stable environment: space observations are free from the turbulence caused by the Earth's atmosphere and very little affected by gravity effects. These unique observing conditions lead to unprecedented high-resolution images and optimal photometric and astrometric accuracies.
- Space observatories offer the unique possibility to observe the same targets for very long periods, which is impossible from the ground for various reasons (day-night or seasonal interruptions, bad weather).
- Finally, observing with a satellite is the only way to have access to the entire sky with the same instrument. This is the guarantee of the homogeneity of the data, essential in many global analyses of the sky and, in the case of astrometry, the only way to obtain absolute measurements of trigonometric parallaxes, then absolute distances.

However, of course, ground-based astronomical observations have many other advantages: the telescopes and their instruments can be constructed in successive steps and progressively improved, they can be repaired which makes their lifetime generally much larger than that of space observatories and new instruments can be installed (in space, this has only been possible with the Hubble Space Telescope, at a very high cost), they can be very heavy and/or very large and, last but not least, they can be operated by astronomers. Finally, ground-based telescopes are less expensive than space telescopes, and the risk associated with building a telescope on the ground is of course much smaller than to launch a satellite.

By *space observatory*, we mean space devices able to globally observe the sky or large parts of the sky, leading to a mass of new information, processed homogeneously. This chapter is concentrating on a few striking examples of such observatories that had, are having, or are expected to have major impacts on our knowledge and understanding of the structure, formation and evolution of the Universe, from the Solar neighborhood to the cosmological background. They are presented by increasing wavelength, from High Energy to Microwave and gravitational wave observatories, through Ultraviolet, Optical and Infrared observatories.

## 2. The Impact of the Earth Atmosphere on Astronomical Observations

The sky as observed from the Earth in the optical (or visible) wavelengths is only a very partial view of all objects observable in the Universe. Indeed, as a function of their temperature, celestial objects emit in various wavelengths, from the extremely high energetic gamma-rays to low radio waves, and different parts of the same object will be scrutinized if observed in different wavelengths. Each part of the whole electromagnetic radiation spectrum will open a new window to the Universe and bring different information. Only from the confrontation of all this information can a consistent picture of the Universe be obtained. Table 1 summarizes key information about the main types of radiation, the typical types of celestial sources observed in these wavelength ranges and examples of space observatories operating in these domains of radiation.

| Type of radiation | Wavelength range | Frequency range (Hz) | Typical sources | Temperature of radiating objects | Examples of space observatories |
|---|---|---|---|---|---|
| Gamma-rays | < 0.01 nm | > $3 \times 10^{19}$ | Compacts objects (from neutron stars to black hole candidates or active galactic nuclei), galaxies, Gamma-Ray bursts. | > $10^8$ K | INTEGRAl, Fermi (ex-GLAST) |
| X-rays | 0.01 – 20 nm | $3 \times 10^{16} - 3 \times 10^{19}$ | Stellar corona, pulsars, star formation regions, colliding galaxies, hot gas in galaxies and clusters of galaxies, supernova remnants, environment of super-massive black holes. | $10^6 - 10^8$ K | Chandra, XXM-Newton, Suzaku *IXO* |
| Ultraviolet | 20- 400 nm | $7.5 \times 10^{14} - 3 \times 10^{16}$ | Very hot stars, supernova remnants, quasars. | $10^5 - 10^6$ K | IUE, FUSE, HST |
| Visible | 400 - 800 nm | $4 \times 10^{14} - 7.5 \times 10^{14}$ | Stars (atmospheres), planets, galaxies, reflection and emission nebulae. | $10^3 - 10^5$ K | HST, Corot, Kepler, Hipparcos, *Gaia* |
| Infrared (IR) | 0.8– 50 mm | $6 \times 10^{12} - 4 \times 10^{14}$ | Cool stars, star forming regions, interstellar dust and gas, planets | $10 - 10^3$ K | ISO, Spitzer, Akari, WISE, *JWST* |
| Far-IR and microwaves | 50 mm - 10 mm | $3 \times 10^{11} - 3 \times 10^{13}$ | Cosmic Microwave Background, cold interstellar medium. | | WMAP, Herschel, Planck |
| Radio | > 1 cm | < $3 \times 10^{11}$ | Interstellar medium, cold molecular clouds, supernova remnants, planets. | < 10 K | - |

Table 1. Wavelengths and type of celestial objects. Right column: examples of satellites within each wavelength range and, in italic, satellites in construction or in project. Adapted from http://outreach.atnf.csiro.au/education/senior/astrophysics/wavebands.html.

All the above is related to the *thermal* radiation emitted by celestial objects considered as blackbodies. As shown in Table 1, the hotter the object, the shorter is the wavelength of the radiation it emits. Some mechanisms also produce *non-thermal* radiation, unrelated to the temperature of the object: synchrotron emission from electrons accelerated or decelerated in a magnetic field (in pulsars or quasars for example); Compton and inverse-Compton scattering increasing or decreasing the energy of X- and gamma-rays, respectively decreasing and increasing their wavelengths (gamma rays from active galaxies, supernova remnants or diffuse gamma rays from molecular clouds; X-rays from accreting black holes or CMB (cosmic microwave background) photons scattered by the electrons in the hot gas surrounding galaxy clusters); masers (microwave-amplified-stimulated emission of radiation) where emission from certain molecular lines can be enormously amplified.

The radiations emitted by celestial objects are very much affected by the Earth atmosphere which is totally or partially opaque to most wavelengths with the notable exception of the optical light and radio wavebands, and those radiations that are not blocked by the atmosphere are suffering various perturbations when crossing it. The Earth atmosphere is a mixture of various gases, mainly Nitrogen ($N_2$), Oxygen ($O_2$), Argon (Ar) and Carbon dioxide ($CO_2$), and water vapor in very small quantities (typically 1 - 4 % close to the Earth), dust, pollen, volcanic ash and other human industrial pollutants. Many other gases are present in extremely small quantities such as Helium (He), Methane (CH4), Hydrogen ($H_2$) or Ozone ($O_3$).

Some of these components have major impacts on the radiations received from celestial objects: most of the infrared, sub-millimeter and microwave radiations are absorbed by water and carbon dioxide molecules, the ultraviolet by ozone and oxygen molecules, the X-ray radiation suffers photo-electric absorption when encountering nitrogen or oxygen atoms in the high atmosphere. The γ-rays up to very high energies are absorbed by atmospheric electrons and nuclei. Ground-based astronomy is again becoming progressively possible for energies above the TeV by the indirect detection of the Cherenkov radiation created by the interaction of high-energy particles (cosmic rays emitted by supernovae explosions, high-energy gamma rays emitted by accreting binary systems, etc.) with the upper atmosphere. The atmosphere is transparent to most of the radio domain except for the shortest wavelengths (below 2 cm) absorbed by water molecules and for the very long ones (larger than a few meters), reflected by the ionosphere back into space.

These effects are illustrated in Figures 2 and 3. Figure 2 gives an overview of the Earth's atmospheric transmittance over the whole electromagnetic spectrum. Figure 3 gives the details of the absorption by several molecules in the ultraviolet, optical and infrared domains.

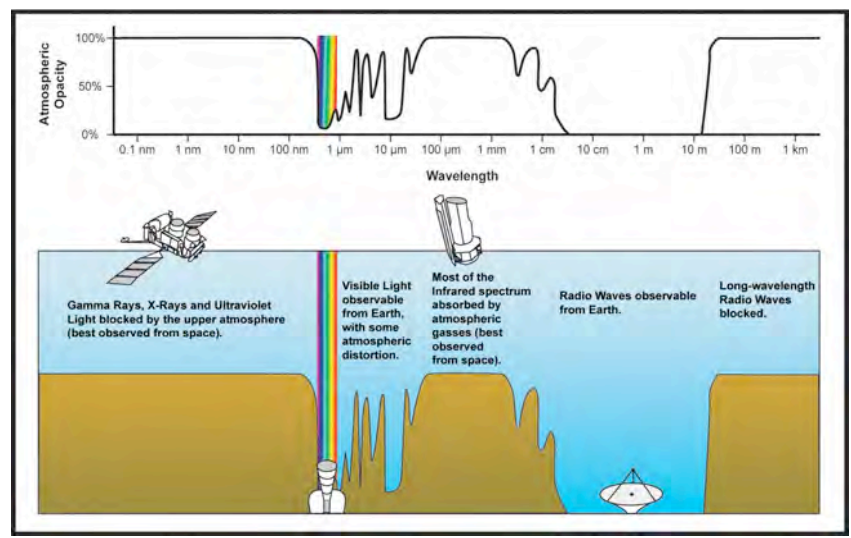

Figure 2. Earth's atmospheric transmittance (or opacity) to various wavelengths of electromagnetic radiation. From
http://coolcosmos.ipac.caltech.edu/cosmic_classroom/multiwavelength_astronomy/multiwavelength_astronomy/orbit.html.

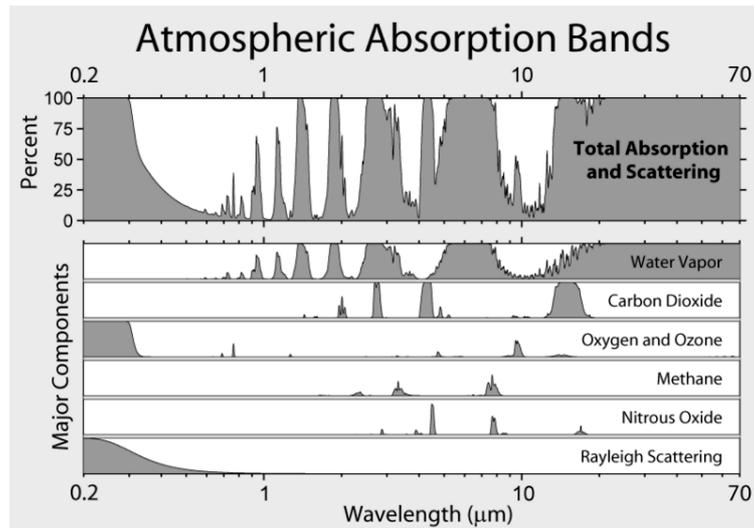

Figure 3. Atmospheric transmission and absorption bands by molecules in the ultraviolet, optical and infrared domains. From http://www.globalwarmingart.com/wiki/
File:Atmospheric_Absorption_Bands_png
(by permission of Robert Rohde).

Besides these absorption effects, the atmosphere also perturbs the radiation that is transmitted to the ground. This explains why there are also satellites observing the sky in visible light. The dust and mist particles in suspension in the atmosphere produce scattering of the light. The visible light is specially affected by scattering as its wavelength is of the same order of magnitude as the diameter of the scattering particles. Also, the atmosphere is constantly in motion and suffers from small variations in temperature and pressure causing motions and distortions to the incoming light. As a result, images of celestial objects are blurred and constantly affected by tiny changes in brightness and position. The apparent position can vary over angular ranges of a few arcseconds. The intensity of these effects in a given location is called *seeing*. The largest telescopes on-ground are situated in high and dry mountains and far from big towns where the seeing is much better. Best sites achieve, rarely, seeing of better than 0.5 arcseconds. Observatories in space do not suffer any distortion from the atmosphere and obtain very stable images and a much better resolution than the best ground-based telescopes, even those obtained with the powerful techniques of adaptive optics that considerably compensate for these seeing effects.

**3. High-Energy Space Observatories**

By the middle of the 20$^{th}$ century, it was known theoretically that a number of different processes occurring in the Universe should produce high-energy photons, such as supernovae explosions or interactions of cosmic rays with interstellar gas. However, since the Earth atmosphere is mostly opaque to high-energy radiation, only observatories situated above it can detect it. For energies greater than about 30 keV, hard (more energetic) X-rays and gamma-rays can be observed from instruments embarked on rockets or balloons, but only satellites, orbiting above the atmosphere, are able to observe the full range of high energies and obtain, through long exposures, enough high-energy photons to achieve detailed studies of the many celestial objects emitting in these wavelengths, thus opening new windows to the unknown. These characteristics explain the very large number of satellites (more than a hundredth) in this domain of energy since the 1960s.

## 3.1. Gamma-Ray Space Observatories

Since 1961 and the first satellite carrying aboard an instrument with a detector designed to detect gamma rays above 50 MeV (Explorer 11), much progress has been made both in the sensitivity of the detectors - increased by factors of 1000 – and in the localization of the sources over the sky, even if this last point remains one of the main difficulties for interpreting of such observations. An unexpected discovery boosted the curiosity of astronomers for this energy domain: the gamma-ray bursts, discovered by American military Vela satellites designed to detect clandestine nuclear bomb tests. Instead of nuclear bomb tests, they detected flashes of inexplicable gamma-ray radiations, and rough estimates for their positions on the sky convinced the military teams that this emission does not originate from either terrestrial or solar sources.

Gamma-ray astronomy explores the most extreme environments and the most violent events in the Universe, places where temperatures can reach hundreds of million degrees, where matter is incredibly dense and/or magnetic fields and gravity extremely strong. Typical targets besides gamma-ray bursts are black holes, neutron stars and pulsars, supernovae, active galaxies and quasars, stellar binaries containing a neutron star or a white dwarf, and many unidentified sources.

**INTEGRAL**
INTEGRAL, INTErnational Gamma-Ray Astrophysics Laboratory, launched on 17 October 2002 from Baikonur on a Russian Proton launcher into a highly elliptical 72 h orbit with an initial perigee of 9000 km and an apogee of 154 000 km, is an ESA-led mission involving Russia, the United States, the Czech Republic and Poland. INTEGRAL remains the most sensitive gamma-ray observatory ever launched and the first space observatory that can simultaneously observe objects in gamma rays, in the range 15 keV to 10 MeV, X-rays and visible light. The satellite is equipped with two gamma-ray instruments: the SPI spectrometer, optimized for high spectral resolution in the energy range 20 keV to 8 MeV, and the IBIS imager, for high spatial resolution providing accuracies in source location of better than 1 arcmin, and two monitors: in X-ray and in the optical V-band.

The first major scientific result from INTEGRAL was the discovery that the low gamma-ray emission from our Galaxy is not due to the interstellar medium but to discrete compact sources, mainly accreting binary systems with a black hole or a neutron star. A second result represents a significant step forward in the identification of the mysterious astrophysical sources that are producing the 511 keV line emission in the centre of the Milky Way, at a rate of about $10^{43}$ positrons per second. This emission of photons is produced via the annihilation of electrons with their antimatter particles, the positrons. INTEGRAL's all-sky map of the 511 keV line emission, obtained via the combination of more than 4 years of observations using the SPI spectrometer (50 million seconds of data!) and an increased spectral and spatial resolution, revealed that the emission is strongly peaked towards the centre of the Galaxy, with an asymmetry along the galactic disc. A careful study of the distribution and characteristics of these 511 keV data led to the conclusion that dark matter is not at the origin of the galactic positron annihilation and that it can be readily explained in terms of classical sources of positrons such as supernova ejecta, winds of Wolf Rayet stars and low-mass X-ray binary systems. The observed asymmetry could be simply explained by the asymmetry of the Galactic spiral arms as seen from the Solar System.

Finally, the publication in July 2010 of the fourth INTEGRAL Soft Gamma-Ray Survey Catalogue, constructed from more than 70 million seconds of observing time with the IBIS imager, now contains more than 700 sources with a substantially increased coverage of extragalactic fields as compared with previous issues of the IBIS survey. The catalogue is now dominated by Active Galactic Nuclei (AGNs, about 30%), followed by High Mass X-Ray Binaries (13%), Low Mass X-Ray Binaries (13%) and Cataclysmic Variables (5%). Unknown sources now constitute nearly 30% of the source list, but the catalogue also provides hard X-ray sources and multi-wavelength follow-up observations that are expected to lead to a large fraction of identifications. A map of the new sources (fourth catalogue compared to the third one) is given in Figure 4, superimposed on the increase in exposure time since the third catalogue.

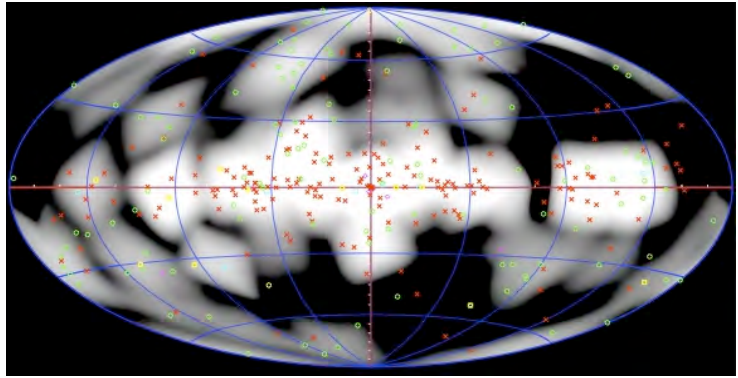

Figure 4. ESA, INTEGRAL mission. Map of incremental exposure since the third catalogue, showing the locations of the new sources found. Key: green circles = AGN; cyan squares = High Mass X-Ray Binaries; magenta diamonds = Low Mass X-Ray Binaries; yellow boxes = Cataclysmic Variables; red crosses = unknown. From http://sci.esa.int/science-e/www/object/index.cfm?fobjectid=45620.
Credit: ESA.

**Swift**
Swift is a NASA-led mission, part of the medium explorer programme (MIDEX), with instruments developed by an international team from the United States, the United Kingdom and Italy, and additional scientific involvement of many other countries. It was launched into a low-Earth circular orbit by a Delta 7320 rocket on 20 November 2004. Swift is a multi-wavelength observatory dedicated to the detection and detailed study of gamma-ray bursts (GRB). Its three instruments are designed to work together to observe GRBs and afterglows in the gamma-ray (Burst Alert Telescope, BAT), X-ray, (X-ray Telescope, XRT), ultraviolet and optical (Ultraviolet/Optical Telescope, UVOT) wavebands. The Burst Alert Telescope is the largest of the instruments, operating in the domain 15-150 keV with a very large field of view (2 steradians) and high sensitivity. Once a GRB is detected, its position is computed on-board with a 3 arc-minute positional uncertainty and relayed within 20 seconds to a network of ground-based observers. In less than about 90 seconds, the spacecraft will autonomously "swift" to that position in order to unable the other two (narrow-field) instruments to observe the burst afterglows and obtain higher accuracy position of the GRB: X-ray spectra between 0.3 and 10 keV will be obtained with the X-ray Telescope, images and spectra (via a grism filter) using the UV/Optical Telescope. For the brightest UV/optical afterglows, the spectra recorded by large ground-based telescopes (such as the VLT) can then be used to determine the redshift of the burst.

The main objectives for the mission are to detect and observe hundreds of GRBs with durations from milliseconds to thousands of seconds and their afterglows, to identify the progenitors of the bursts and characterize their host galaxies, to perform detailed multi-wavelength analysis of their afterglows in order to better understand the evolution of the blast and its interactions with the surroundings, and finally to use these observations to better understand the early Universe.

There is a huge variety of gamma-ray bursts: some of them only last fractions of second while others last a few hundred seconds, occasionally longer. They are followed by *afterglows* in longer wavelengths, observed over much longer times. A major breakthrough in the observation of these extreme phenomena came from the Italian-Dutch satellite Beppo-SAX in 1997, which provided the first precise burst position and discovered its X-ray afterglow, providing the confirmation of the cosmological distances of the bursts. This discovery opened the way to the discovery of afterglows in the whole radiation domain and to their systematic observation. The emerging interpretation of these observations is that the shortest bursts are emitted when two compact objects - either a pair of neutron stars or a neutron star and a black hole - collide and merge, and the origin of longer bursts is the collapse and death of (super) massive stars exploding as "hypernovae".

Soon after its launch, on 27 December 2004, Swift was one of many observatories that observed the brightest gamma-ray source ever seen from outside the solar system: an exceptionally bright flare from the galactic SGR 1806-20, a strongly magnetized neutron star called magnestar, about 15 000 pc away, in the Sagittarius constellation. The total flare energy was about a hundred times higher than the other two previously observed giant flares. The tremendous luminosity of the initial gamma-ray spike was consistent with the catastrophic release of (nearly) pure magnetic energy from the magnestar and probably occurred during a catastrophic reconfiguration of the neutron star's magnetic field.

On 13 April 2010, Swift discovered its 500$^{th}$ burst, GRB 100413B, a long burst in the constellation Cassiopeia. An all-sky map showing the locations of the 500 first gamma-ray bursts detected by Swift is given in Figure 5. These GRBs may be as close as about 30 million pc and as far away as 4 billion pc, covering a span of time equivalent to about 95 percent of the universe's age. The farthest burst of this impressive series was observed on 23 April 2009: GRB 090423 in Leo, 4.3 billion pcs away, is one of the most distant objects known in the Universe.

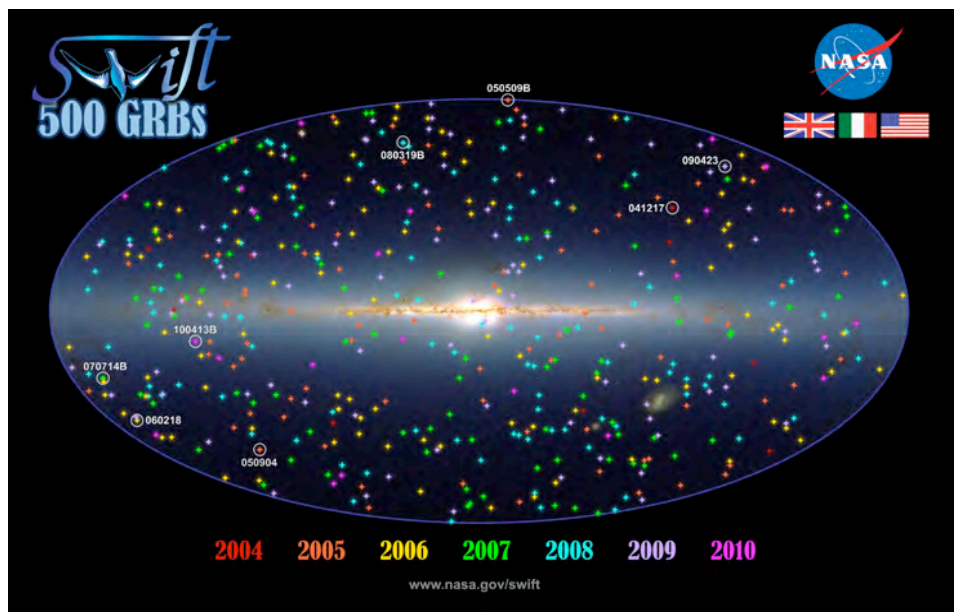

Figure 5. NASA. Swift mission. All-sky map showing the locations of Swift's 500 first gamma-ray bursts, color coded by the year in which they occurred. In the background, an infrared image shows the location of our galaxy and its largest satellites. From http://www.nasa.gov/mission_pages/swift/bursts/500th.html.
Credit: NASA/Swift/Francis Reddy.

**Fermi**

The NASA-led Fermi Gamma-ray Space Telescope (formerly GLAST) is the last in the series of gamma-ray observatories: it was launched on 11 June 2008 from Cape Canaveral by a Delta II rocket in a low-Earth circular orbit and is the result of an international and multi-agency cooperation between NASA, the U.S. Department of Energy and institutions in France, Germany, Japan, Italy and Sweden. It is operated in the energy range 10 keV - 300 GeV and is also aiming at the exploration of most extreme environments in the Universe. Its main instrument, the Large Area Telescope (LAT), is a gamma-ray, large field of view, imager (20 MeV - 300 GeV) providing all-sky coverage several times per day. Operating more like a particle detector, the LAT uses 880000 silicon strips within a 1.8-meter cube, and is able to detect high-energy gamma rays with unprecedented resolution and sensitivity: it is about 30 times more sensitive than any previous satellite gamma-ray detector and is able to determine the location of a source to better than half an arc-minute. The second instrument, the Gamma-ray Burst Monitor (GBM), is an all-sky monitor in the range 10 keV - 25 MeV aiming at the detection of transient phenomena such as occultations and gamma-ray bursts.

Published in May 2010, the first Fermi-LAT catalogue is the result of the first 11 months of data. It includes 1451 sources, from star burst galaxies and active galactic nuclei (AGN) to galactic pulsars, supernova remnants (SNR), X-ray binary stars (HXB) and micro-quasars. 630 of the sources remain unidentified, not associated with sources detected at lower energies. A map showing the sky distribution of the whole catalogue in galactic coordinates is given in Figure 6.

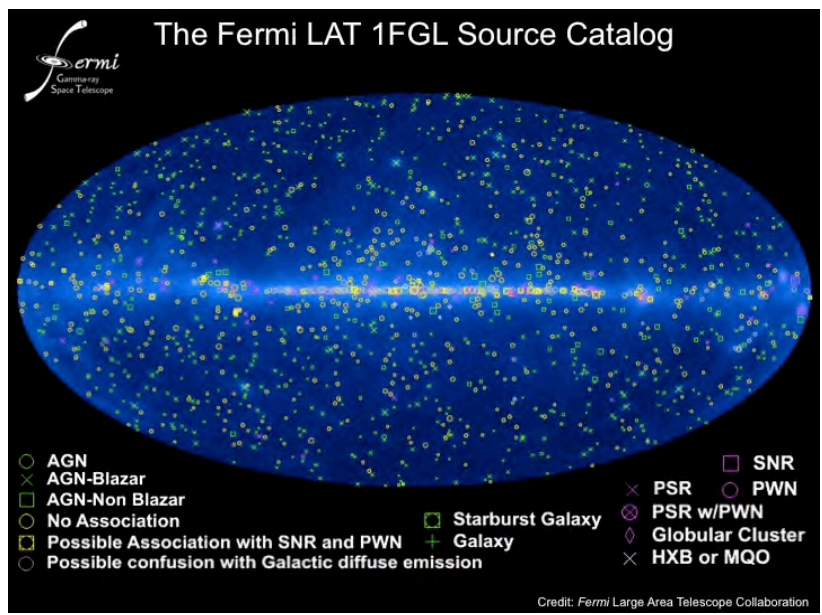

Figure 6. Fermi Gamma-ray Space Telescope's first all-sky catalogue, obtained from 11 months of sky survey data using Fermi's Large Area Telescope (LAT). AGN = active galactic nuclei, PSR = pulsars, PWN = pulsar wind nebulae, SNR = supernova remnants, HXB = x-ray binary stars, MQO = micro-quasars. From http://apod.nasa.gov/apod/ap100318.html. Credit: NASA, DOE, International Fermi LAT Collaboration

One remarkable highlight of Fermi is the observation of Centaurus A, one of the closest radiogalaxies, situated at distance of about 4 million pc. In its centre there is a super-massive, very active, black hole, from where jets of magnetized particles are ejected, producing strong emission at many different wavelengths. A radio-optical-gamma-ray composite image of the galaxy is shown in Figure 7. The optical image of the host giant elliptical galaxy, NGC 5128, is in the centre. The diffuse high-energy gamma radiation detected by Fermi's LAT is the much larger purple halo, and the full extent of Centaurus A is given by the giant radio-emitting lobes (color-coded in orange), stretching to more than 0.4 Mpc. This diffuse gamma-ray emission is explained by the collision between the highest-energy particles of the radio lobes and the cosmic microwave background photons. This process was known to produce X-rays in many active galaxies, but it is the first time that microwave photons are shown to be up-scattered to gamma-ray energies.

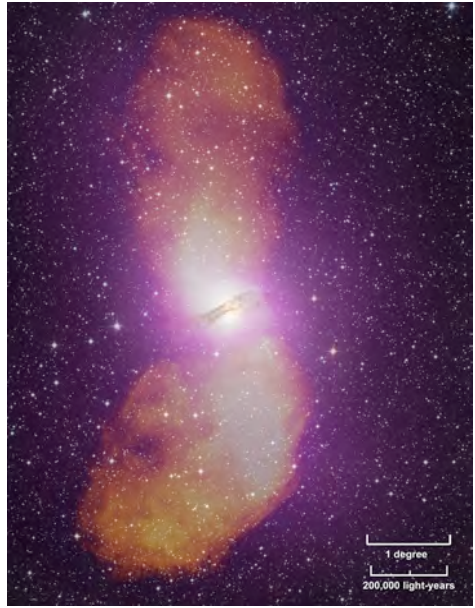

Figure 7. Radio/optical/gamma-ray composite image of the active radiogalaxy Centaurus A. In the centre: the optical image. In purple: the diffuse gamma-ray emission observed by Fermi's LAT. In orange: the radio emission. From http://www.nasa.gov/mission_pages/GLAST/news/smokestack-plumes.html.
Credit: NASA/DOE/Fermi LAT Collaboration, Capella Observatory, and Ilana Feain, Tim Cornwell, and Ron Ekers (CSIRO/ATNF), R. Morganti (ASTRON), and N. Junkes (MPIfR)

**3.2. X-Ray Space Observatories**

X-rays are the signature of the hot Universe: from the solar corona to hot gas in the most distant clusters of galaxies, from planets to stars and supernova remnants, from hot gas in star forming regions to colliding galaxies or gamma-ray bursts afterglow, X-rays are an extremely powerful tool to explore the Universe at all scales. Figure 8 is an illustration of the major differences between the X-ray sky and the optical sky. The image on the left shows the constellations of Orion and Canis Majoris with the Moon to the top. To the right is the same area of the sky as imaged in X-rays. Sirius is visible to the bottom left of each image. However, in the optical it is Sirius A, the brightest star in the visible sky, and in X-rays it is Sirius B, the white dwarf companion of Sirius A. The bright blue source to the top of the X-ray image is the Crab Pulsar surrounded by the Crab nebula while the Moon is very faint.

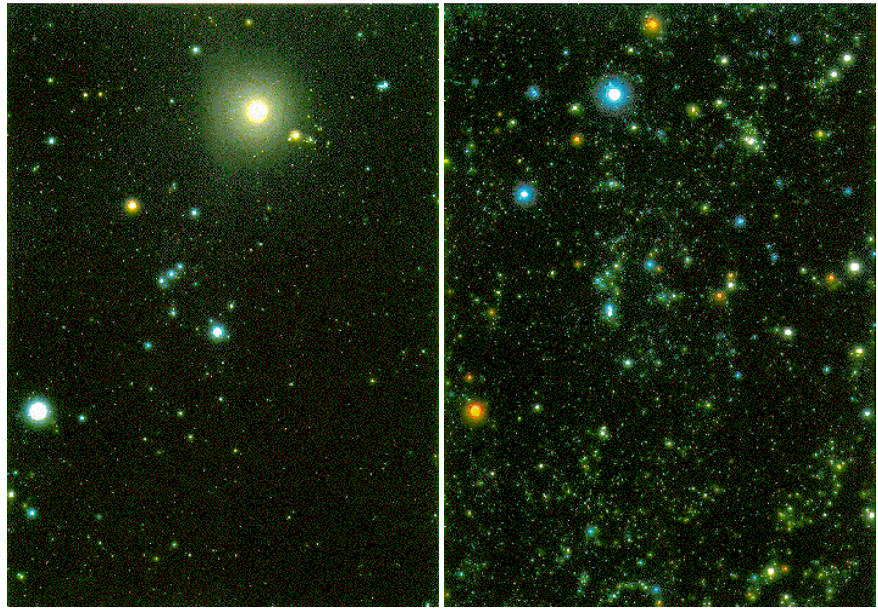

Figure 8. The X-ray sky as compared to the optical sky. These are two images of the constellations of Orion and Canis Majoris. Left: in optical wavelengths. Right image: in X-rays, showing the hottest objects of the field. From http://www-xray.ast.cam.ac.uk/xray_introduction/History.html: The History of X-ray Astronomy, Cambridge X-Ray Astronomy Group.

The atmosphere is totally opaque to X-rays and, from the 1960's, many X-ray instruments have been used to study the X-ray Universe from space. The first attempts were just detectors carried to the upper atmosphere on rockets. An X-ray test detector on board a V2 rocket, launched on 5 August 1948, recorded the first solar X-ray. Then a few other attempts were made on the Sun showing that the level of emission from the Sun was so low that it was not worth trying to observe other celestial objects. However, in 1962, a detector with a much-improved sensitivity was embarked on a rocket to try and detect the reflected X-ray emission from the Moon. It indeed detected the emission from the Sun and the Moon, but also, unexpectedly, a strong emission from the Scorpius constellation: a first X-ray source was discovered outside the Solar System: Scorpius X-1. This discovery opened the way to a new field of astronomy: the exploration of the Universe in X-ray band. Riccardo Giacconi received the Nobel Prize in Physics in 2002 for developing this new research domain.

From the 1960's, much progress has been made - and is still being made – in X-ray technology, especially in the detectors and in the X-ray optics (much lighter mirror assembly, then much larger collecting areas, better angular and spectral resolution), greatly increasing the overall sensitivity and the ability to focus X-ray radiations. These advances are allowing high-quality images and spectra and lead to detailed studies of millions of X-ray sources. Several X-ray satellites are now in operation: RXTE, the Rossi X-ray Timing Explorer (launched in 1995) and Chandra (launched in 1999) operated by NASA, XMM-Newton (launched in 1999) operated by ESA, and Suzaku, the fifth Japanese X-ray astronomy satellite (launched in 2005). The next generation of X-ray observatories is an international cooperative project, pursued by ESA, NASA and JAXA (the Japanese space Agency): the International X-ray Observatory (IXO). It builds on three decades of successful X-ray telescope development and is candidate as one of the next big missions of the three space agencies.

**X-ray telescopes**
X-ray telescopes are very different from optical telescopes. Because of their high-energy, X-ray photons directly arriving on a mirror would not be reflected by it but would penetrate into it. The most commonly used technique is to make X-rays ricochet off grazing-incidence mirrors which are nested in a coaxial and cofocal configuration. A schematic illustration of grazing incidence in X-ray telescopes is shown in Figure 9 (top left) with only four nested pairs of mirrors. The grazing angles range from about 3.5 degrees for the outer pair to about 2 degrees for the inner pair. In Figure 9 (bottom left) is shown a cutaway of the design and functioning of the High Resolution Mirror Assembly on Chandra. The mirrors are coated with a highly reflective rare metal, the iridium. A photo of one of the three Mirror Modules of XMM-Newton is given in the right of Figure 9. Each Mirror Module consists of 58 gold-coated nested mirrors. Each mirror shell consists of a paraboloid and an associated hyperboloid, precisely aligned. The thickness of the smallest mirror (diameter=306 mm) is 0.47 mm, and it increases linearly with shell diameter in order to guarantee sufficient stiffness. The thickness of the 700 mm diameter mirror is 1.07 mm. The minimum radial separation between adjacent shells is 1mm. Indeed, in grazing incidence optics the effective area of a telescope is a function of the number of mirrors, the mirror shell thickness and their separation. The thinner the mirror shells are and the narrower the shells are spaced, the larger is the collecting area.

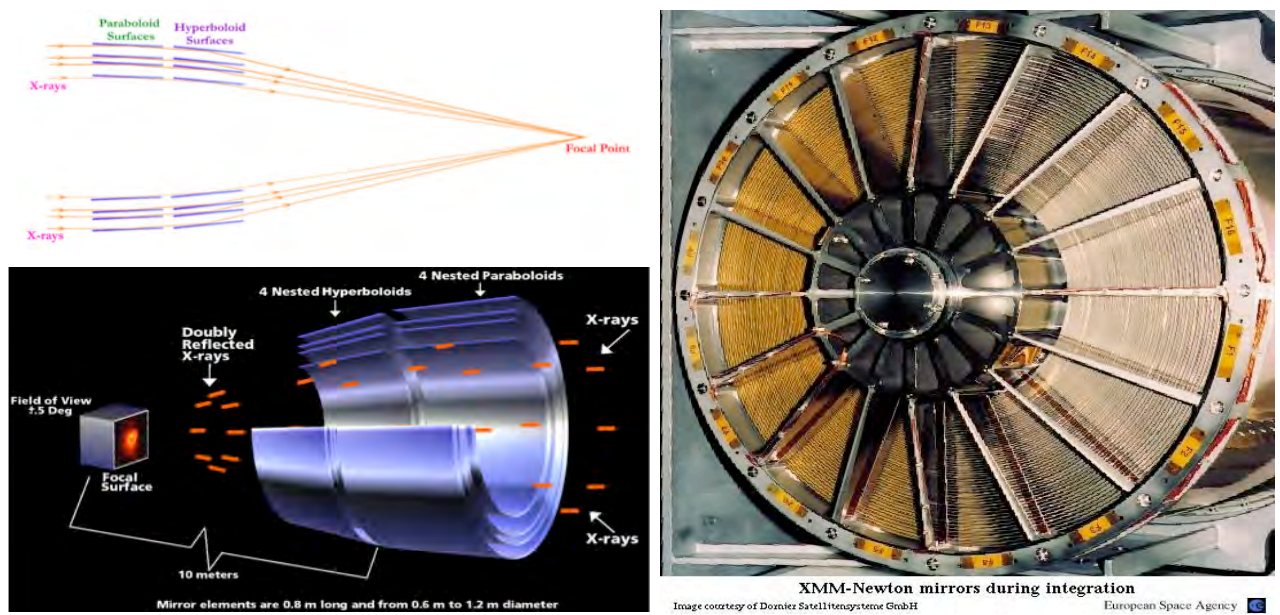

Figure 9. Top left: X-ray optics: Principle of grazing incidence reflection and focusing of X-rays (Credit**:** NASA/CXC/S. Lee). Bottom left: Schematic design of the High Resolution Mirror Assembly on Chandra (Credit**:** NASA/CXC/D. Berry).
From http://chandra.harvard.edu/resources/illustrations/teleSchem.html#xray_mirror.
Right: One the three XMM-Newton mirror modules with a full set of 58 flight mirror nested shells. From http://xmm.esac.esa.int/external/xmm_science/gallery/public/level3.php?id=65.
Credit**:** Dornier Satellitensysteme GmbH and ESA.

**Chandra**
NASA's Chandra X-ray Observatory was launched and deployed by Space Shuttle Columbia on 23 July 1999 in a high elliptical 64-hour orbit (16 000 to 133 000 km). Such an orbit allows uninterrupted observations as long as 55 hours. Chandra is designed to observe X-rays in the wavelength range 0.1 to 10 keV. Its instruments have approximately fifty times better spatial resolution than the previous big X-ray observatory, ROSAT (the ROentgen SATellite, 1990-1999, Germany, USA, and UK).

This is illustrated in Figure 10 with images of the Crab nebula: the image taken by the Advanced CCD Imaging Spectrometer (ACIS) on Chandra shows how higher resolution can reveal important new features.

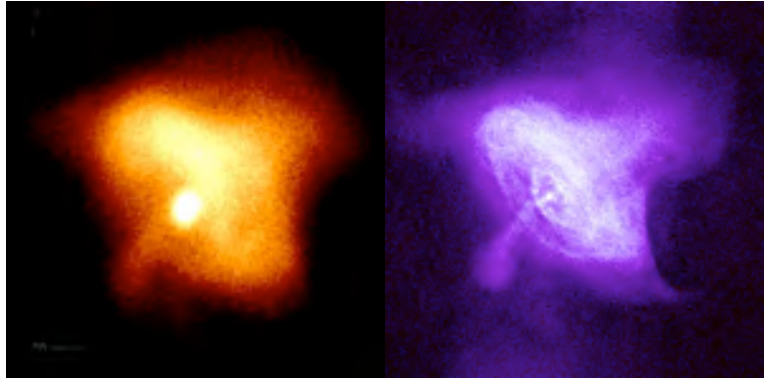

Figure 10. An illustration of the progress of spatial resolution in X-ray observations: the Crab Nebula observed by ROSAT (left, credit: S. L. Snowden, USRA, NASA/GSFC) and by Chandra (right, credit: NASA/CXC/SAO). From http://chandra.harvard.edu/about/axaf_mission.html.

This Chandra image of the Crab Nebula and its pulsar was one of the early observations obtained with the telescope and it is a striking illustration of the observatory's capability for high-resolution imaging of violent phenomena. This image led to a major discovery: the bright inner elliptical ring in the Nebula was showing the first evidence of the shock front where the wind of particles from the pulsar begins to radiate in X-rays via the synchrotron process.

Chandra combines the observing possibilities of four different instruments: two cameras, the High Resolution Camera (HRC) and the Advanced CCD Imaging Spectrometer (ACIS), and two high-resolution spectrometers, the High Energy Transmission Grating Spectrometer (HETGS) in the energy range 0.4 to 10 keV and the Low Energy Transmission Grating Spectrometer (LETGS) in the energy range of 0.08 to 2 keV. HRC is the camera used to identify the fainter sources and make high-resolution (0.5 arcsecond) images of areas full of hot matter, for example supernova remnants or clusters of galaxies. ACIS can make images in a very narrow range of energy centered on the lines produced by specific ions (for example oxygen, neon or iron ions) and therefore, by taking different images, study temperature and/or chemical variations across clouds of hot gas. The spectrometers are used in the study of detailed energy spectra measuring temperature, ionization and chemical composition of the observed targets.

Chandra's capability for high-resolution imaging is enabling detailed mapping of the structure of extended X-ray sources and its high angular resolution permits studies of faint discrete sources. Also important is its contribution to high-resolution dispersive spectroscopy.

Chandra's high-resolution imaging capability could be illustrated by many spectacular images. Two of them are given in Figures 11 and 12. Figure 11 is an image obtained from a 164-hour exposure (11 pointings over nearly three years) of the centre of our own galaxy. It shows Sagittarius A*, with more than 2000 other X-ray sources and a diffuse extended emission of hot, X-ray-emitting gas, heated and chemically enriched by numerous stellar explosions. Figure 12 presents a composite image of the nearby galaxy NGC 7793, combining X-rays observations made with Chandra, optical data from the ESO's Very Large Telescope and H-alpha data from the Cerro Tololo Inter-American Observatory 1.5m telescope. This image shows, in the outskirts of the galaxy, a microquasar containing a black hole with the most powerful jets ever seen from such a stellar-mass black hole.

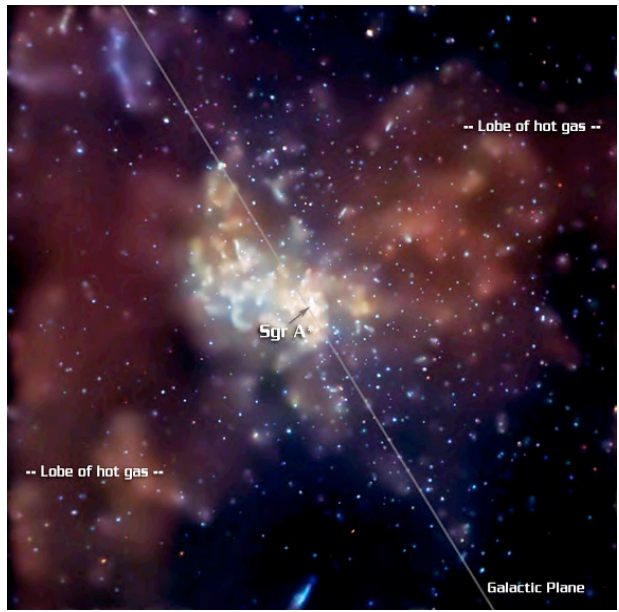

Figure 11. Chandra image of the supermassive black hole at the centre of our Galaxy. The locations of Sagittarius A* and of the galactic plane are indicated.
From http://chandra.harvard.edu/photo/2003/0203long/. Credit: NASA/CXC/MIT/F.K.Baganoff et al.

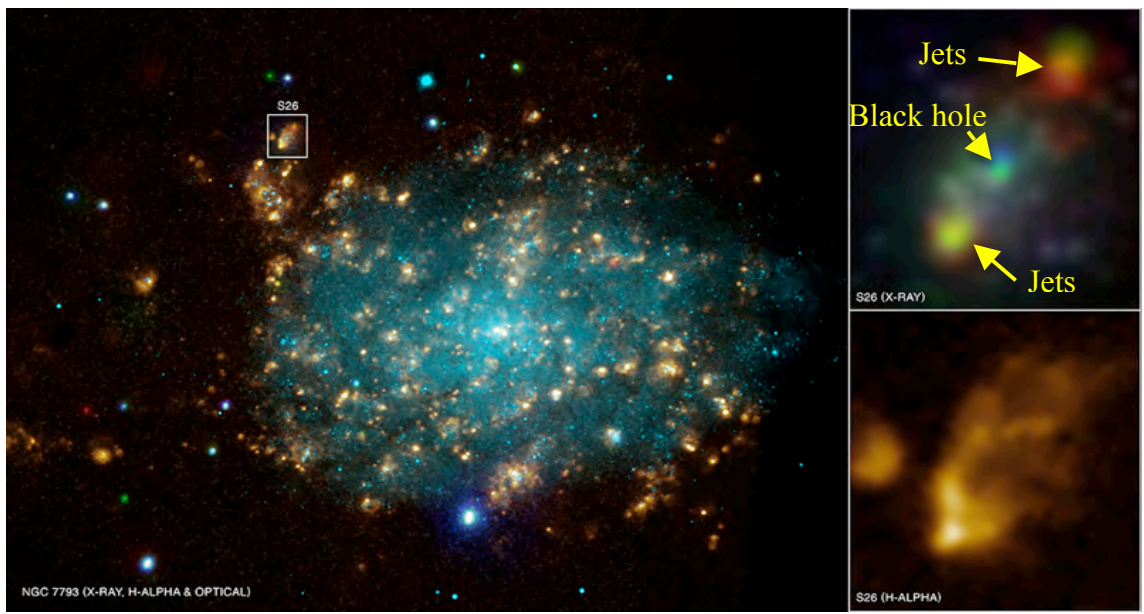

Figure 12. Composite image of nearby galaxy NGC 7793 combining X-ray observations obtained with Chandra's ACIS instrument (red: 0.2-1.0 keV, green: 1.0-2.0 keV, blue: 2.0-8.0 keV), optical data from the ESO's Very Large Telescope (cyan) and H-alpha data from the CTIO 1.5-m telescope (gold). The wide field image of the whole galaxy (left) is 9 arcmin across (about 10 000 pcs); the detailed images of the microquasar (right) in X-rays and H-alpha are 45 arcsec wide (about 900 pcs).
From http://chandra.harvard.edu/photo/2010/ngc7793/. Credits: X-ray image: NASA/CXC/Univ of Strasbourg/M. Pakull et al.; Optical image: ESO/VLT/Univ of Strasbourg/M. Pakull et al.; H-alpha image: NOAO/AURA/NSF/CTIO 1.5m.

**XMM-Newton**

XMM-Newton was launched by Ariane-5 from Kourou, French Guiana, on 10 December 1999 and placed into a 48-hour elliptical orbit around the Earth. The ESA's X-ray Multi-Mirror (XMM) observatory, renamed XMM-Newton short after launch, is a powerful soft X-ray observatory, concentrating on the radiation range 0.1 – 10 keV. With a perigee altitude of 7000 km and an apogee at 114 000 km, the satellite is traveling out to nearly one third of the distance to the Moon. This very eccentric orbit enables very long and uninterrupted observations. The satellite is the second cornerstone of ESA's Horizon 2000 space science plan, devoted to High-Throughput X-ray Spectroscopy. With three large mirror modules (see Figure 9), three European Photon Imaging Cameras (EPIC) measuring the proportions of different X-ray wavelengths, two Reflection Grating Spectrometers (RGS) diffracting the X-rays to achieve high spectral resolving power (150 to 800) over a wavelength range from 5 to 35 Å (0.33 to 2.5 keV), and an optical-UV monitor, XMM-Newton's capabilities are very complementary to those of NASA's Chandra. As said in a recent review on the first decade of science with Chandra and XMM-Newton (Santos-Lleo et al. 2009): "The complementary capabilities of these observatories allow us to make high-resolution images and precisely measure the energy of cosmic X-rays. Less than 50 years after the first detection of an extrasolar X-ray source, these observatories have achieved an increase in sensitivity comparable to going from naked-eye observations to the most powerful optical telescopes over the past 400 years". As a result of this increase in sensitivity, the number of observed objects also has tremendously increased. This is illustrated in Figure 13, showing the increase in size of X-ray source catalogues produced over the past four decades. XMM-Newton is making about 40 000 new detections per year. A vast majority of these sources (98%) had never been detected before in X-rays.

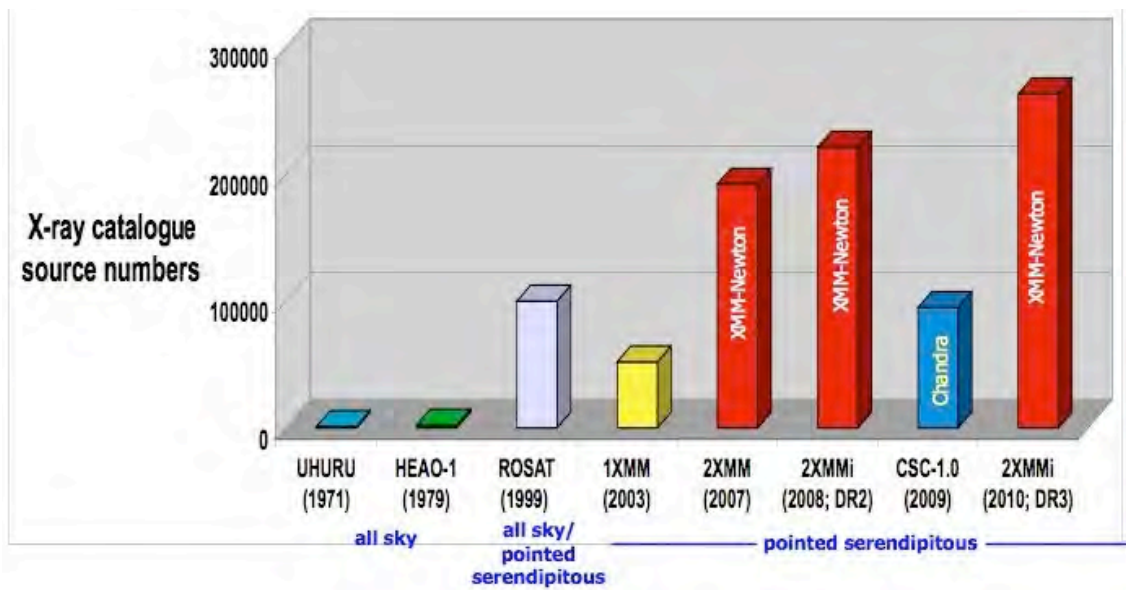

Figure 13. This diagram illustrates the increase in size of X-ray source catalogues that have been produced over the past four decades. From left to right, they are UHURU (1971), HEAO-1 (1979), ROSAT (1999), 1XMM (2003, First XMM-Newton Serendipitous Source Catalogue), 2XMM (2007, Second XMM-Newton Serendipitous Source Catalogue), 2XMMi (2008; DR2, Incremental Second XMM-Newton Serendipitous Source Catalogue, second data release), CSC-1.0 (2009, First release of Chandra Source Catalog) and the latest release of the second XMM-Newton Serendipitous Source Catalogue, 2XMMi-DR3 (2010; DR3).
From http://sci.esa.int/science-e/www/object/index.cfm?fobjectid=46961.
By permission of M. Watson, University of Leicester.

Figure 14 provides one example of the discoveries made possible by XMM-Newton: a huge cloud of high-temperature gas in the Orion Nebula. The cloud is composed of winds blowing from high-mass stars that are heated to millions of degrees. This result suggests that such X-ray cloud should be common in star-forming regions.

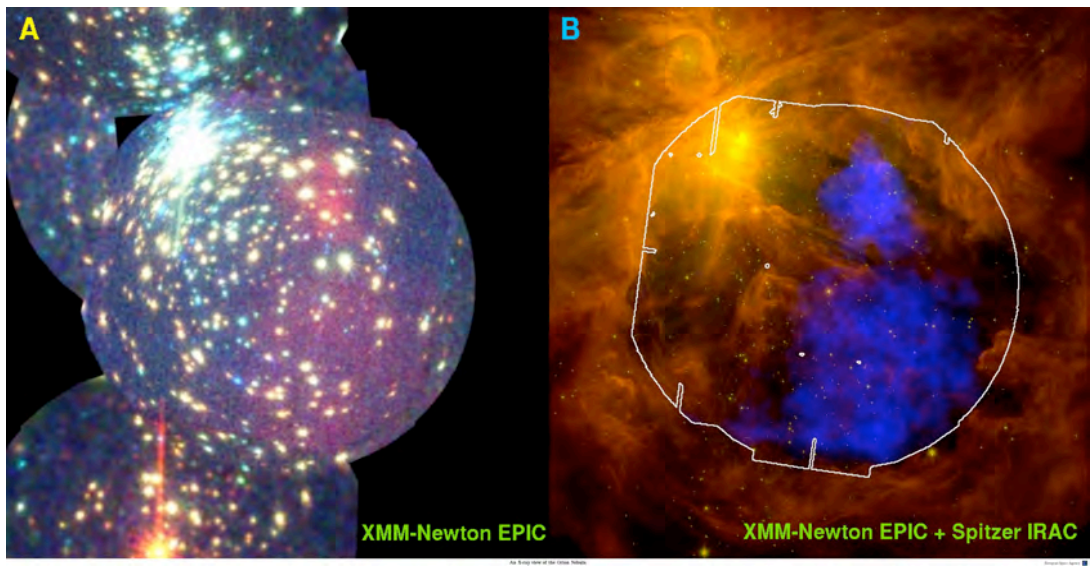

Figure 14. The Orion Nebula. The left panel is an X-ray image obtained with ESA's XMM-Newton, with the hot gas seen as a red haze. The right panel, a Spitzer (NASA) infrared image of the Orion Nebula overlaid with XMM-Newton X-ray data (in blue), shows the newly discovered hot gas cloud. From http://sci.esa.int/science-e/www/object/index.cfm?fobjectid=46060. Credit: XMM-Newton EPIC, Guedel et al. (left); AAAS/Science, ESA XMM-Newton and NASA Spitzer data (right).

Another very striking result is the contribution of XMM-Newton to the COSMOS survey (Cosmological Evolution Survey), designed to probe the formation and evolution of galaxies as a function of cosmic time (redshift) and large-scale structure environment. Covering a 2 square degree equatorial field, it is a collaboration between XMM-Newton and Chandra (X-rays), the Hubble Space Telescope (optical), Spitzer (infrared), GALEX (UV) and a number of large ground-based telescopes. Over 2 million galaxies have been detected, spanning 75% of the age of the Universe. The X-ray emission as observed by XMM-Newton is shown in Figure 15. One of the goals is to establish the link between X-ray emission and underlying dark matter from the study of one of the largest samples of X-ray-detected clusters of galaxies.

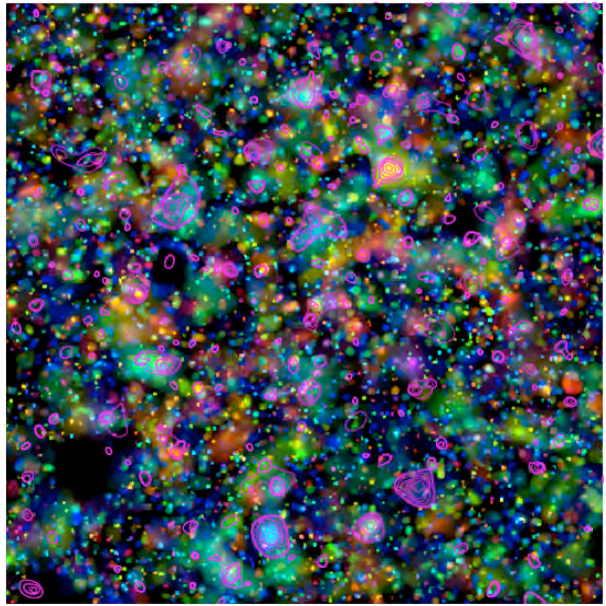

Figure 15. X-ray emission in the COSMOS field. This image shows the galaxy density with colors representing the redshift of the galaxies ranging from redshift of 0.2 (in blue) to 1 (in red). The X-ray contours (in pink) show the extended X-ray emission as observed by XMM-Newton. From http://sci.esa.int/science-e/www/object/index.cfm?fobjectid=46328. Credit ESA.

**IXO**

IXO, the International X-ray Observatory, is the next generation project of X-ray space observatory, developed jointly by ESA, NASA and JAXA (the Japan Space Agency) for a launch in 2020 at the earliest. It is the result of the merger, in 2008, of two preliminary mission concepts developed by ESA (XEUS = X-Ray Evolving Universe Spectrometer) and NASA (Constellation X). It is building on the accumulated scientific and technologic knowledge acquired by the current and previous X-ray facilities. With a large X-ray mirror (about 3 square meter collecting area and 5 arcsec angular resolution), it is planned to observe the hot Universe between 0.1 and 40 keV with a powerful suite of instruments able to deliver high-resolution spectroscopy, deep spectral detailed imaging over a wide field of view, microsecond spectroscopic timing with high count rate capability and unprecedented polarimetric sensitivity. Its main science drivers are the formation and evolution of galaxies, clusters and large-scale structures (co-evolution of galaxies and their central supermassive black holes, creation of chemical elements, chemical evolution along cosmic time), black holes and matter under extreme conditions of gravity, temperature, magnetic field, and the life cycles of matter and energy.

## 4. Optical-Ultraviolet Space Observatories

### 4.1. Ultraviolet Space Observatories

The atmosphere is opaque to most of the ultraviolet radiation, except for near UV that is only observable in high mountain observatories. UV radiation ranges approximately from 10-20nm (extreme UV) to 380-400nm (near UV) and is emitted by hot gas and dust grains in the interstellar medium, in the atmosphere of young massive stars or very old stars, in novae, interacting binary stellar systems, debris of supernovae, AGN accretion discs, and can be used as a tracer of star formation and galactic evolution.

**IUE**

Among the few UV space observatories, IUE, the International Ultraviolet Explorer satellite (NASA-ESA-UK) was one of the world's longest space mission (nearly 19 years of orbital operation, from January 1978 to September 1996) and the first general user UV space observatory. IUE was a 45-cm telescope equipped with two on-board spectrographs covering ultraviolet wavelengths from about 120 to 340 nm. Its operation was very flexible (less than one hour to point at a new target) and it was used nearly as a ground-based telescope. About 30000 spectra have been obtained, from about 9000 targets extending from comets to quasars. Among the many results obtained from these data, let us quote the first identification of the progenitor of any supernova in history (Supernova 1987A), the first discovery of high velocity winds in stars other than the Sun or the first direct determination of the size of the active regions in the nuclei of Seyfert galaxies (mini-quasars).

**EUVE**

EUVE, the NASA Extreme Ultraviolet Explorer, was designed to operate in the relatively unexplored extreme ultraviolet range of the spectrum, from 7 - 76 nm. It was long thought that this radiation would be totally blocked by the interstellar medium, absorbed by Hydrogen. This was contradicted in 1975 with the discovery of the strong extreme UV emission from a hot white dwarf. This discovery and results showing the patchiness of the local interstellar gas opened the way to EUVE. It was launched in June 1992 and operated until January 2001. Its primary objective during the first six months following launch was to explore this new window: the satellite carried out a full-sky survey in this extreme ultraviolet band and simultaneously performed a deep survey along a narrow band on the Ecliptic. These surveys led to the detection of about 1000 sources (illustrated in Figure 16), including a wide variety of astronomical objects: hot white dwarfs, active and nearby late-type stars, cataclysmic variables, and various types of active galactic nuclei. EUVE made the link between X-ray and far-UV observations.

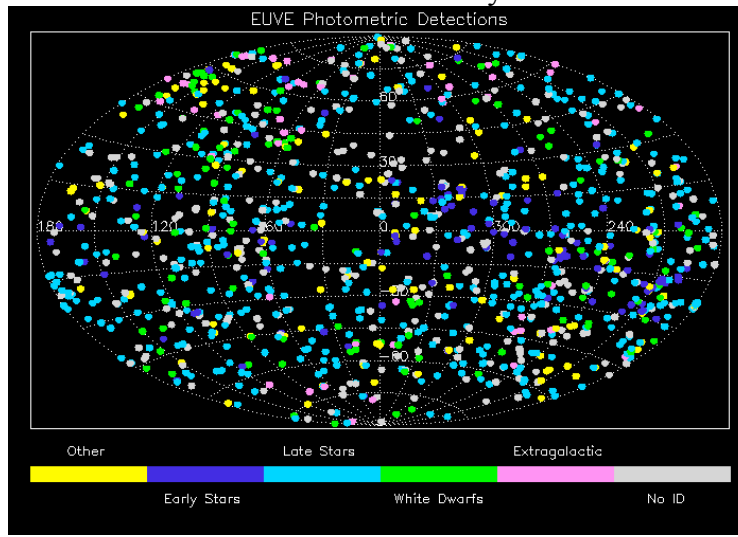

Figure 16. EUVE Photometric detections in the extreme UV radiation domain. From http://archive.stsci.edu/euve/allsky/results.html. Credit: NASA, C. Damian.

**FUSE**

FUSE, Far Ultraviolet Spectroscopic Explorer, was a NASA mission, with cooperation from the Canadian Space Agency (CSA) and the French Space Agency (CNES), aiming at high-resolution spectroscopy in the far-ultraviolet spectral region, from approximately 90 to 120 nm. It was operated between June 1999 and October 2007, performing about 6000 observations of nearly 3000 sources. FUSE was especially investigating the tenuous regions of planetary atmospheres, stellar envelopes, interstellar and intergalactic gas in the Solar System, our Milky Way and nearby galaxies, and even distant active galaxies and quasars. FUSE data were, for example, used to search for various molecules (deuterium, molecular hydrogen, etc.) and ionized gas (OI, OVI, CIV, etc.) in the interstellar medium near the Sun in the Local Cavity, in the Milky Way disc, in the Magellanic System and further, showing the presence of hot gas in the Galactic halo and of a very tenuous corona extending further than the halo, into the nearby intergalactic medium. FUSE also provided new insights in the study of gas in proto-planetary and debris discs, showing for example a carbon overabundance in the debris gas orbiting Beta Pictoris.

**GALEX**

The Galaxy Evolution Explorer (GALEX) is a NASA Small Explorer Class mission that is investigating star formation in galaxies over 80% of the history of the Universe. GALEX is being operated since April 2003. The GALEX instruments allow imaging and spectroscopic observations to be made in two ultraviolet bands, far UV (135-178 nm) and near UV (177-273 nm), with spatial resolution 4.3 and 5.3 arcseconds respectively. The satellite is performing a series of nested imaging and spectroscopic surveys to characterize star formation in the local Universe and investigate the evolution of star formation over cosmic time, from the early Universe up to the present, with the aim of understanding the major phenomena driving star formation. With a field of view of 1.2 square degrees, GALEX performed an all-sky imaging survey, deep sky surveys in the imaging mode, and partial surveys in the near and far UV spectroscopic modes, and mapped tens of millions of galaxies. In such a massive database, galaxies of all ages and status are found. Two examples of intermediate age spiral galaxies with vigorous star formation are shown in Figure 17: NGC 300 (left) and NGC 1291 (right). NGC 1291 has a very unusual inner bar and a beautiful outer ring structure.

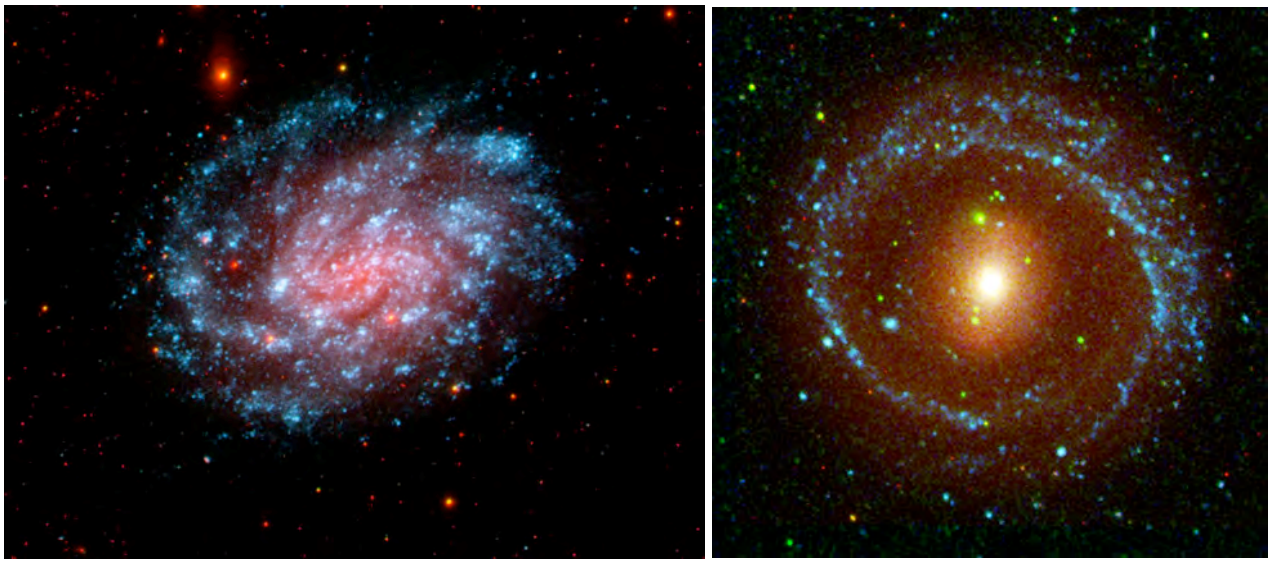

Figure 17. GALEX observation of two intermediate age spiral galaxies. Left: NGC 300. Right: NGC 1291. Color codes: blue = GALEX near UV, green = GALEX far UV, red = visible light from the Las Campanas Observatory, Chile (for NGC 300) or from the Cerro Tololo Inter-American Observatory, Chile.
NGC 300: From http://www.jpl.nasa.gov/news/features.cfm?feature=1524. Credit: NASA/JPL-Caltech/Las Campanas.
NGC 1291: From http://photojournal.jpl.nasa.gov/catalog/PIA10114. Credit: NASA/JPL-Caltech/CTIO

**Hubble Space Telescope**

Finally, three instruments of the Hubble Space Telescope are also designed to observe in the UV wavelengths: the Cosmic Origins Spectrograph (COS), with two detectors covering the domain 115 - 320 nm, the Space Telescope Imaging Spectrograph (STIS), which covers a very wide range of wavelengths from the near-infrared region (1 μm) into the ultraviolet (115 nm), and the Advanced Camera for Surveys (ACS) observing from the far-ultraviolet, through the visible and out to the near-infrared (120 to 1000 nm). They are described in the following section.

## 4.2. Optical Space Observatories

While the Earth atmosphere is transparent to visible wavelengths, there are a number of problems with ground-based observations which are making optical space observatories more powerful than ground-based telescopes in many respects. The first one is turbulence. The light has to travel through successive turbulent layers that are blurring the images of celestial objects before reaching telescopes on the ground. During the last decades, methods have been developed to overcome this problem: speckle interferometry, limited to very bright objects, and adaptive optics which is now used by most large ground-based telescopes. However, adaptive optics is currently mainly limited to the near-infrared wavelength range and to small fields of view. Also, it is mainly designed to observe moderately extended objects and to work with the assistance of guide stars, which limits its use to the vicinity of suitable, bright enough, objects. For telescopes operated from above the atmosphere on a satellite, there is no such turbulence effect. As a result, an observatory such as the Hubble Space Telescope can produce high-resolution razor-sharp images.

Also, unlike a ground-based telescope, a space observatory can reach every part of the sky and does not suffer daily or seasonal interruptions, offering the possibility of full-sky surveys. It can operate twenty-four hours a day and all year round and, if required, make extremely long exposures. Other Earth perturbations on instruments such as gravitational flexure, thermal distortions, wind shaking, etc. can be obviated or controlled in space, allowing a very high stability of the images. Thanks to these very stable observing conditions and to the absence of turbulence, very high accuracies and very low systematic errors can be reached on photometric, astrometric or shape measurements. This is one of the major rational for satellites in essentially three domains: space astrometry, detection of transits of exoplanets and stellar seismology, and all-sky mapping of the distortions of galaxy images to measure the weak lensing effects produced by dark matter.

### 4.2.1. Hubble Space Telescope

The Hubble Space Telescope (HST) is a joint NASA-ESA project. It was launched on 25 April 1990 by the Space Shuttle Discovery (STS-31) into a circular orbit, described in 96 minutes, 569 km above the ground. With a 2.4-m primary mirror and five instruments, it is designed to take high-resolution images and accurate spectra across the entire optical spectrum, from the near infrared, through the visible, to the far-ultraviolet light. The telescope has been designed to be a permanent space-based observatory that could be regularly repaired and upgraded by astronauts, and indeed, five "servicing missions" have taken place between December 1993 with Space Shuttle Endeavour and recently in May 2009 with Atlantis. A spectacular view of the telescope over Earth, taken on 19 May 2009 by the crew of the Space Shuttle Atlantis, is shown in Figure 18.

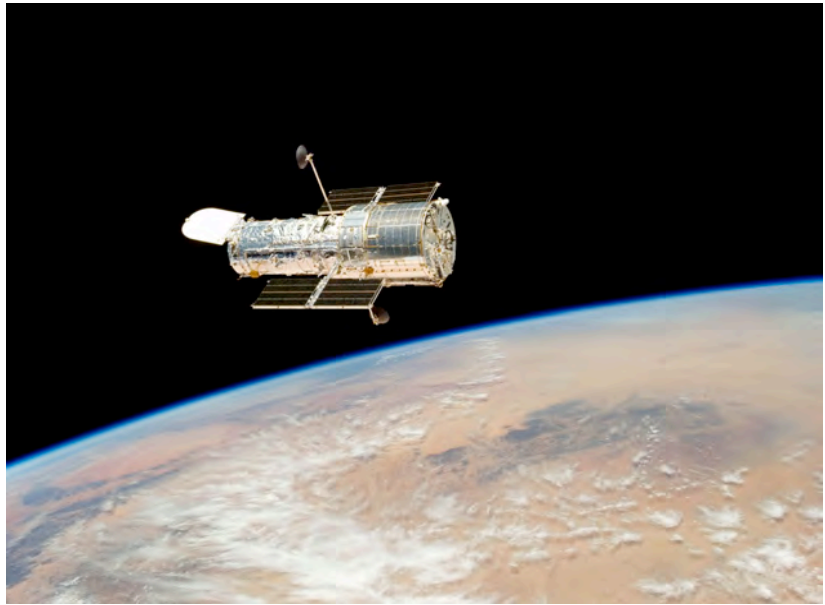

Figure 18. Hubble over Earth after its release on 19 May 2009. Photo taken by the crew of the Space Shuttle Atlantis after the very successful Servicing Mission 4.
From http://www.spacetelescope.org/images/heic0908c/. Credit: NASA/ESA.

From the last servicing mission in May 2009, HST is equipped with five high performance science instruments including two cameras and three imaging spectrographs, and with the three fine guidance sensors. These instruments can work either together or individually. They are housed in the bottom third of the telescope. A schematic projected drawing of the position of the five instruments and of the three Fine Guidance Sensors is shown in Figure 19 and an exploded view of the whole telescope in Figure 20. The characteristics of each of the instruments are given below.

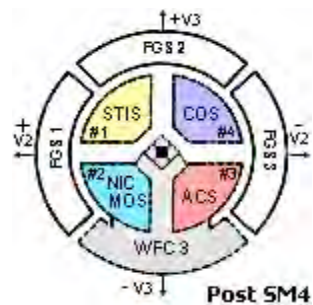

Figure 19. Schematic drawing of the position of the five instruments and of the three Fine Guidance Sensors. From http://hubble.nasa.gov/technology/instruments.php. Credit: NASA.

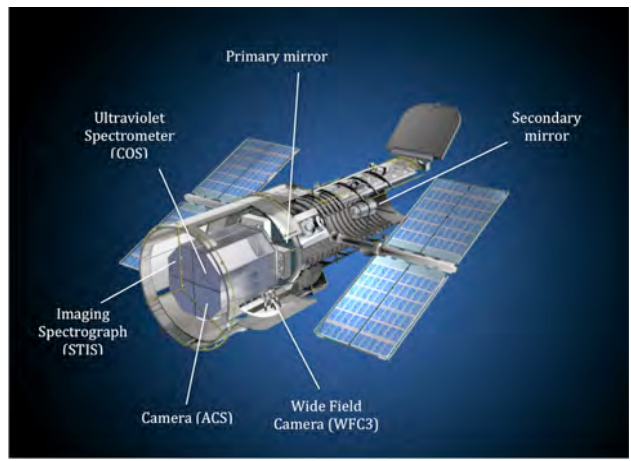

Figure 20. View of Hubble showing the instruments and the whole system. Adapted from http://www.spacetelescope.org/images/exploded_view/. Credit: ESO.

The Wide Field Camera 3 (WFC3) is a fourth-generation instrument installed in the Hubble Space Telescope during Servicing Mission 4 (SM4). It offers improved high-resolution imaging over a wide field of view (2.7x2.7 arcmin in the visible channel) with a large wavelength range in two channels, one for far-ultraviolet and visible light (200-1000 nm) and the other one for near-infrared (850-1700 nm). It offers comparable performance to the ACS instrument, but over a wider range of wavelengths.

The Cosmic Origins Spectrograph (COS) is a fourth-generation instrument installed during SM4. COS has two 2.5 arcsec diameter circular apertures and is designed to perform high sensitivity, medium- and low-resolution spectroscopy (medium-resolution about 20000, low-resolution about 2500) of astronomical objects in the ultraviolet. It covers ultraviolet wavelengths from the far-ultraviolet (115 to 177 nm) to the near-ultraviolet (170-320 nm). COS complements the possibilities of the STIS spectrograph instrument (larger field of view and much larger wavelength range).

The Advanced Camera for Surveys (ACS) is a third-generation instrument that replaced Hubble's Faint Object Camera during Servicing Mission 3B. Its wavelength range extends from the far-ultraviolet, through the visible and out to the near-infrared (120 to 1000 nm), and its name comes from its wide field of view (3.4x3.4 arcmin) and particular ability to map large areas of the sky in great detail. It includes a sub-instrument, the High Resolution Camera, designed to take extremely detailed pictures in a small field of view (29x26 arcsec) and equipped with a coronagraph improving the HST contrast near bright objects.

The Space Telescope Imaging Spectrograph (STIS) was installed during SM2, stopped functioning in August 2004 due to a power supply failure, and was repaired during SM4. STIS is sensitive to a wide range of wavelengths, from the ultraviolet to the near-infrared (115 to 1000 nm), and is able to produce the spectrum of spatially extended objects, such as galaxies, covering many points across the image simultaneously. STIS also has a coronagraph able to block light from bright objects and make possible investigations of nearby fainter objects.

The Near Infrared Camera and Multi-Object Spectrometer (NICMOS) was installed during SM2 and repaired during SM3B. It is providing infrared imaging and spectroscopic observations within the near-infrared wavelength range, from 0.8 to 2.5 μm. Its field of view varies from 11x11 arcsec at high resolution to 51x51 arcsec at low resolution. Since late-2008, the instrument is experiencing difficulties with its cooling system and it was not available for observations in 2010.

The three Fine Guidance Sensors (FGS) are optical sensors (467-700 nm). Two of them are used for pointing the telescope with high accuracy and with a 0.007 arcsec stability over long period of time for about 95% of the time. The third one is used for relative astrometry. The FGS have been refurbished or replaced several times (SM2, SM3A and SM4).

The Hubble Space Telescope is world-wide known for its dramatic and so aesthetic images of a huge variety of celestial objects, but it is also, with its complete set of very performing instruments, a major tool to investigate the Universe in detail, from the Solar System to the early Universe. A few examples are presented here, selected among the most recent highlights.

Star and planetary formation have been studied in very different environments from the closest nebulae in our own Galaxy to very distant galaxies. Within 1 kpc from the Sun, the Orion nebula was the object of several extensive studies using all imaging instruments aboard the telescope. More than 3000 stars of various sizes were observed, from the largest ones, blowing winds of dust and gas, to the smallest brown dwarfs. Hubble mosaic image of the region shows the arcs, blobs, pillars, and rings of dust and gas created from the material ejected by the most massive stars and gives clues to the history of star formation in this patchy area. Further on, an HST *Treasury Program* obtained deep multi-color photometry and detailed pictures of thousands of sources and provided, in December 2009, an atlas of protoplanetary discs formed around newly formed stars. It was shown that discs formed close to bright stars are different from those formed farther away, providing a characterization of the dust grains possibly at the origin of planetary formation. Star formation was also studied in globular clusters, showing evidence of several successive generations of stars, or in the Magellanic Clouds where three successive generations of stars have created shells of gas and dust further and further away from the centre of the observed massive nebula. Images of a stellar nursery in the Carina Nebula, situated about 2300 pcs away, is presented in Figure 21, showing the very different aspects of this turbulent area of stellar formation when observed in the visible wavelengths or in the near infrared. The visible light emphasizes the streams of charged particles blown out from the very hot newly formed stars while a multitude of stars, hidden by gas and dust in the visible image, are observed in the infrared image.

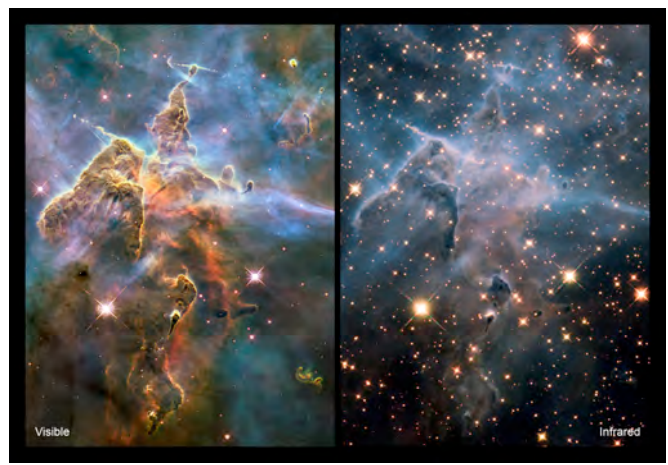

Figure 21. Images of a stellar nursery in the Carina Nebula taken with the WFC3 installed in May 2009. Left: visible-light image. Right: near-infrared image. From
http://www.spacetelescope.org/images/heic1007b/.
Credit: NASA, ESA, M. Livio and the Hubble 20th Anniversary Team (STScI).

One of the first images obtained by the Wide Field Camera 3 (WFC3) installed on board Hubble in May 2009 is the Butterfly planetary nebula (NGC 6302) shown in Figure 22, left. The gas, heated to more than 20 000 K, is ejected at more than 950 000 kilometers per hour by an exploding star and is forming a bright *planetary nebula*. The image reveals the complex history of the successive gas ejections, and the variety of filters available in WFC3 provides information about the temperature, density and composition of the gas. Another important field of activity is the study of galaxies, and Hubble showed the stunning variety of such objects in the Universe. One example is the well-known and photogenic Whirlpool galaxy. An HST composite image in visible wavelengths is given in Figure 22 (right). It shows the hydrogen emission associated with the most luminous young stars in the spiral arms of the galaxy.

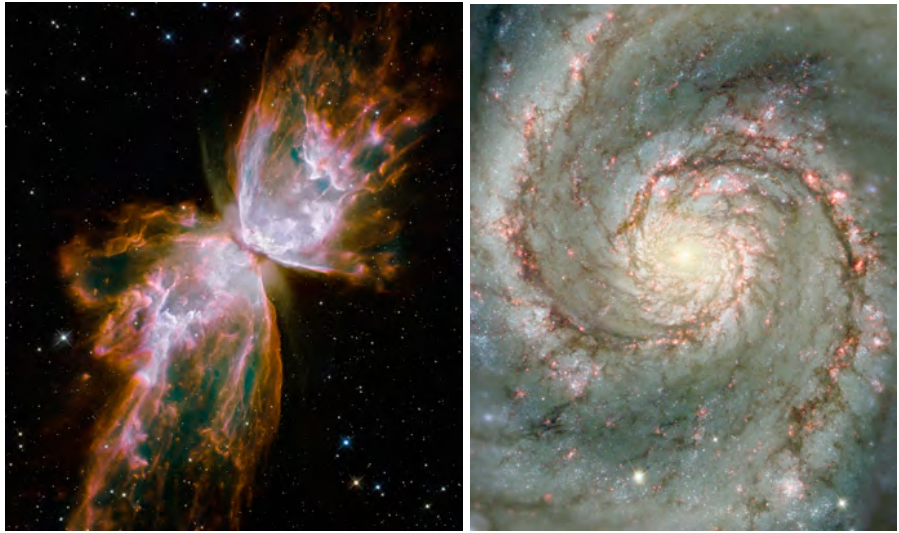

Figure 22: Left : The Butterfly planetary nebula NGC 6302, observed with the WFC3 installed in May 2009. From http://www.spacetelescope.org/images/heic0910h/. Credit NASA, ESA and the Hubble SM4 ERO Team
Right: The Whirlpool galaxy, M51, observed with the Wide Field Planetary Camera 2. From http://www.spacetelescope.org/images/opo0110a/. Credit NASA/ESA and the Hubble Heritage Team STSCI/AURA

Hubble also made images of the deep Universe. One example is the galaxy cluster Abell 370 observed with Advanced Camera for Surveys, repaired during SM4. Figure 23 (left) shows a composite image of the cluster made using filters that isolate light from green, red and infrared wavelengths. It illustrates the phenomenon of gravitational lensing resulting in arcs and elongated images. These distorted images are caused by the cluster's gravitational field that twists the light from galaxies lying far behind it. The mass distribution in the galaxy cluster, including the dark matter, can be reconstructed from the observations of the gravitationally lensed images of background galaxies. A second example is one of the deepest view of the Universe ever done: the *Hubble Ultra Deep Field* (HUDF), obtained from 800 exposures taken with ACS over 400 Hubble orbits i.e. a total exposure time of 11.3 days (Figure 23, right). The image includes nearly 10 000 galaxies, mainly dwarf galaxies, very few bright ones. This is interpreted as the evidence that the galaxies are building up from the merging of smaller ones, as predicted by the hierarchical theory of galaxy formation. A joint detailed analysis of these data and of many complementary data obtained by Chandra and XMM-Newton in the X-rays, by Spitzer in the infrared, by Hershel in the far-infrared, and by many large ground-based telescopes, also shows that the star formation rate in these dwarf galaxies was about ten times higher than is happening now in nearby galaxies.

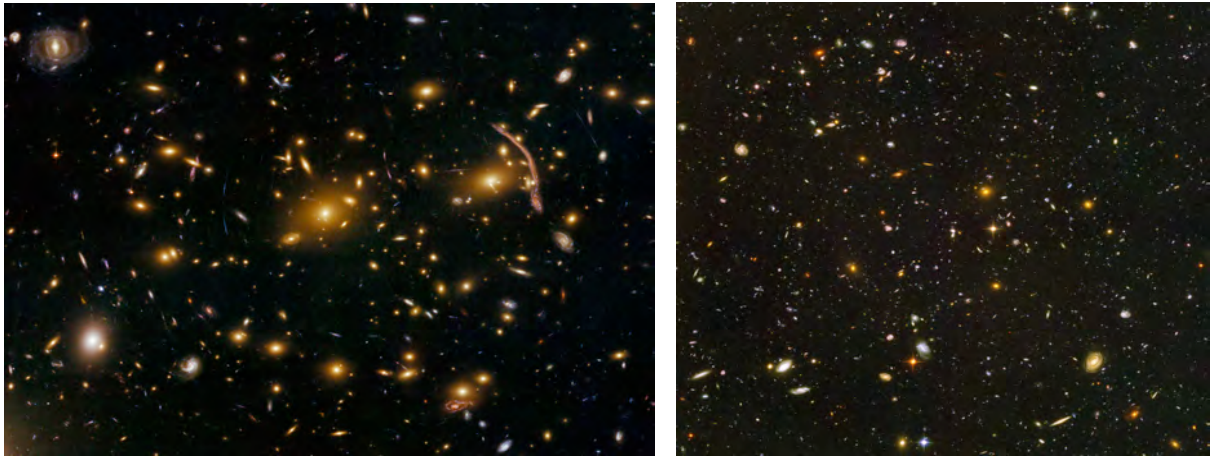

Figure 23. Left: Gravitational lensing in the galaxy cluster Abell 370 taken with the ACS. From http://www.spacetelescope.org/images/heic0910b/. Credit NASA, ESA, the Hubble SM4 ERO Team and ST-ECF. Right: The Hubble Ultra Deep Field taken with the ACS. From http://www.spacetelescope.org/images/heic0611b/. Credit: NASA, ESA, and S. Beckwith (STScI) and the HUDF Team.

### 4.2.2. Astrometry from Space

The determination of the distances, and motions, of celestial objects is one of the major keys of the understanding of the Universe and many efforts have been put in the derivation of methods to measure or estimate these. Most of these methods ultimately rely on the precise distance determination of the nearest stars, subsequently used as calibrators for distance determination of more distant objects. Indeed, it is only for the nearest stars that it is possible to detect and measure the tiny angular variations produced by the motion of the Earth around the Sun (the trigonometric parallaxes), and to disentangle them from the two other major sources of angular displacements over the celestial sphere: the motions of the stars with respect to the Sun (the proper motions), and the orbital motions caused by the presence of one or more companions (star or planet) orbiting the observed star.

From ground, positional measurements are running into many problems: turbulence of the Earth atmosphere, thermal and gravitational distortions of the instruments and inability of ground-based observatories to observe the whole celestial sphere. These problems were strongly hampering any potential progress in ground-based astrometry: by the 1970s, the average error on trigonometric parallaxes was just below 0.01 arcsec with unexplained North-South systematic differences of up to 0.005 arcsec, and the difficulty to obtain reliable trigonometric parallaxes from ground was such that only a few tens of them were published each year. One of the key input to modern astrophysics, the accurate distance of stars – opening the way to their intrinsic luminosity and to the 3-D structure of our environment – was then out of reach.

As early as 1965, only seven years after the launch of the first Sputnik, a nonconformist French astrometrist, P. Lacroute from Strasbourg Observatory, proposed to go to space to perform astrometric observations with a much higher accuracy. The basic principles of global space astrometry that he proposed were subsequently adopted for Hipparcos and now for Gaia: a revolving satellite with two apertures providing two simultaneous viewing directions separated by a large, fixed, angle (58° for Hipparcos; 106.5° for Gaia). This concept is permitting direct angular measurements between stars situated in far away areas of the sky, thus directly providing *absolute* trigonometric parallaxes (unlike ground-based or Hubble observations, restricted to small field *relative* observations).

**Hipparcos**

The first space astrometry satellite, Hipparcos (HIgh Precision PARallax Collecting Satellite), was included in the ESA's Science Programme in 1980, launched by Ariane from Kourou in August 1989 and operated until 1993. It was a modest-size telescope with a mirror of 29-cm diameter. The light from the stars in the two fields of view was focused on and modulated by a grid situated in the focal plane and containing 3000 parallel slits. It was then registered by an Image Dissector Tube, precisely positioned on each of the 118000 pre-selected stars. The Hipparcos Catalogue, published by ESA in 1997, provides very accurate and homogeneously measured astrometric parameters for the 118000 program stars (down to magnitude 12, but mainly brighter than magnitude 10). Median standard errors in parallax and annual proper motion are in the range 0.7-0.9 milliarcsec (mas) for stars brighter than magnitude 9. A deeper study of micro-perturbations of the satellite attitude due for example to micro-meteorites or thermal effects on the solar panels led to a new reduction of the Hipparcos raw data published in 2007, with a large improvement of the accuracies of the astrometric parameters for the brightest stars (average parallax error of 0.3 mas for stars brighter than magnitude 7). In addition to the Hipparcos Catalogue obtained from the main detection device of the satellite, the measurements primarily used for the satellite attitude reconstruction have also been used for a systematic astrometric and photometric cartography down to magnitude 12: the Tycho Catalogue provides two-color photometry and lower accuracy astrometric parameters for one million stars, rather complete to about $V = 10$. It was superseded in 2000 by the Tycho-2 Catalogue, including two million stars and improved proper motions obtained from early epoch ground-based positions combined to Tycho positions.

The Hipparcos results, of unprecedented accuracy and homogeneity, are impacting a very broad range of topics: from the provision of an accurate reference frame, allowing the rigorous re-reduction of historical astrometric catalogues and improved astrometric data for recent surveys, to a precise understanding of the structure, kinematics and dynamics of the Solar neighborhood, through many applications in stellar structure and evolution thanks to an accurate dataset of fundamental parameters for a variety of stellar types. Indeed, with some 22000 stars for which distances were determined to better than 10% (as compared to a couple of hundreds for ground-based measurements, mainly for fainter stars) and 50000 to better than 20%, our knowledge of the stellar populations in the Solar neighborhood has been greatly enhanced. By mid-2010, more than 2500 refereed papers (6000 if all bibliographic sources are considered) are explicitly using the Hipparcos or Tycho Catalogues for applications as diverse as the measurement of the General Relativistic Light bending, element diffusion and convective motions in stellar interiors, 3-D structure and kinematics of nearby clusters and associations, determination of the galactic rotation parameters, evidence for halo accretion, determination of the ages of galactic open and globular clusters, mass determination of asteroids, Earth rotation or age of the Universe. Besides astrometry, the very homogeneous Hipparcos photometry (about 110 observations per star all over the mission in the Hp broadband) also provides an accurate multi-epoch database widely used for variability studies. Only three examples will be given below.

The Hipparcos HR diagram, obtained for the 16 631 single stars with a relative error on trigonometric parallax smaller than 10% and an error on $B-V$ smaller than 0.025 magnitude is presented in Figure 24. For the first time for field stars, the clump of red giants was very prominently visible, thus providing a very precise calibration of their mean absolute magnitude, thereafter used as absolute distance indicator within our Galaxy and outside.

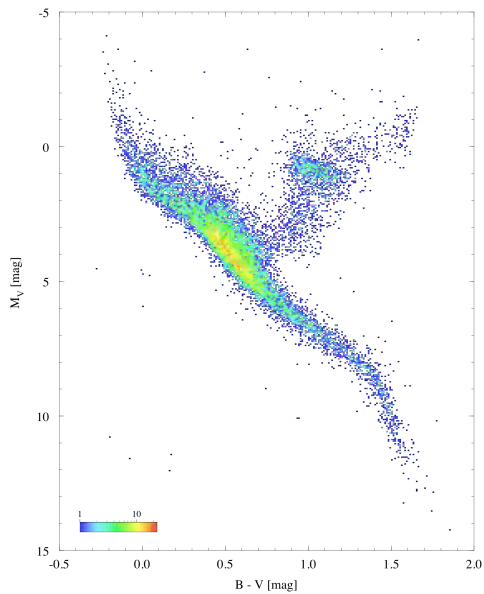

Figure 24. Hipparcos HR diagram: absolute magnitudes $M_V$ versus color index $B-V$ for the 16 631 single stars with a relative error on trigonometric parallax smaller than 10% and an error on $B-V$ smaller than 0.025 magnitude. From
http://www.rssd.esa.int/index.php?project=HIPPARCOS&page=HR_dia. Credit ESA

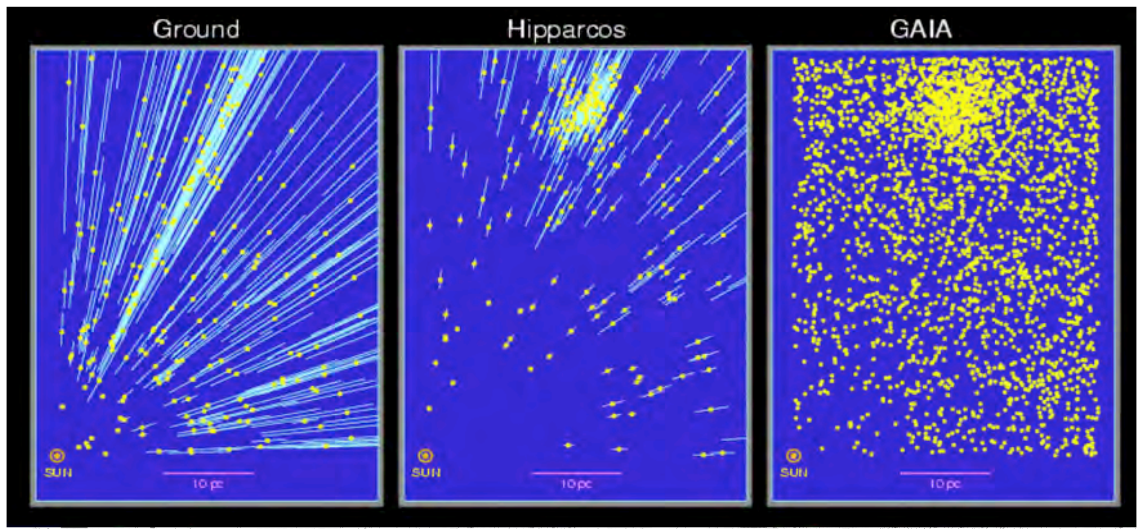

Figure 25. Trigonometric parallaxes of stars in the direction of the Hyades cluster. Left: from ground-based parallaxes as available in 1970. Middle: from the Hipparcos Catalogue (1997). Right: as expected from Gaia observations (simulation). In each of the drawing, all stars are included at the distance from the Sun calculated from the trigonometric parallax (yellow dots) and the blue lines are showing the uncertainties in the determination of the distance. In the right panel, the uncertainties are so small that they are not visible at the scale of the drawing. From
http://www.rssd.esa.int/index.php?project=GAIA&page=IG_Accuracy_2.
Credit: ESA, M.A.C. Perryman.

Hipparcos data have been extensively used for studying open and globular galactic clusters, dynamical streams and moving groups, and associations of young stars. Besides the discovery of new clusters, membership probabilities have been greatly improved, individual distances of star members and the 3-D structure and kinematics of the nearest clusters have been determined, leading to the characterization of individual members and to the cluster's age. These results provide the basis for further studies of stellar structure, effect of stellar rotation, constraints on models of stellar evolution, relation age-chemical composition of clusters, etc. The trigonometric parallaxes of stars in the direction of the Hyades cluster as available from ground-based observations in the 1970s (left), from the Hipparcos Catalogue (middle), and as expected from the coming generation of space astrometry satellite, Gaia (right) are shown in Figure 25. The 3-D structure of the Hyades cluster is already clearly shown in the middle panel while Gaia will provide an extremely detailed description of the cluster and of its environment.

A comprehensive census of OB associations within 1 kpc from the Sun was made from Hipparcos positions, proper motions and parallaxes determination leading to a major improvement in membership probabilities, a much better knowledge of the relation of associations to nearby molecular clouds, and significantly narrowing the locus of stars in the color-absolute magnitude diagram, with a clear single-star main sequence. A panorama of nearby OB-associations is shown in Figure 26.

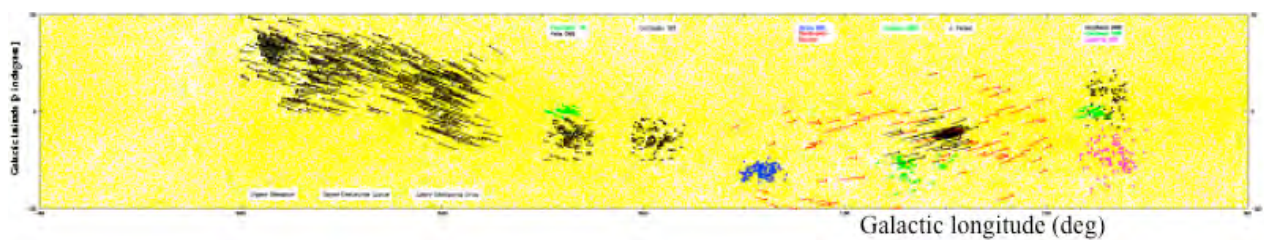

Figure 26. Panorama of nearby OB-associations observed by Hipparcos. Credit: ESA, J. de Bruijne.

The success of Hipparcos led to the definition of a much more ambitious follow-up mission: Gaia, included in the ESA's Science Programme in 2000 and planned for launch by the end of 2012. Built on the same principle as Hipparcos, a revolving satellite with two widely separated fields of view, it benefits from modern technologies (especially for the detectors and the mirrors) and is aiming at the systematic observation of a billion objects all over the sky down to magnitude 20. Moreover, in addition to astrometric observations of exquisite accuracies (from 7 micro-arcseconds for the brightest part of the program to 300 micro-arcseconds for the faintest), dedicated on-board instruments will provide spectrophotometry in the red and blue wavelength ranges for all objects observed in astrometry, and spectroscopy for stars brighter than magnitude 17. These two instruments will provide complementary radial velocity data and astrophysical diagnostics. Gaia will provide an unprecedented stereoscopic survey of our Milky Way and the nearby universe including 1 billion stars, 300 000 solar system objects, millions of galaxies, 500 000 quasars, 10 000 exoplanets. It will be the astronomical data archive for decades to come, with a tremendous discovery potential when combined with other archives. An artist view of the satellite is shown in Figure 27.

Prospects for space astrometry after Gaia are of two different types: all-sky observations with micro-satellites such as Micro-JASMINE (Japan) or JMAPS (USNO, USA) or more topical satellites aiming at still much more accurate measurements of the galactic bulge and parts of the plane (JASMINE, observing in the near-infrared – Japan) or the detection of habitable planets with astrometric accuracies down to 1 micro-arcsec (SIM Lite – NASA-JPL). Astrometry had to wait until the end of the $20^{th}$ century to be accurate enough to become an essential tool for astrophysics and it is now opening many new ways of exploration for the $21^{st}$ century.

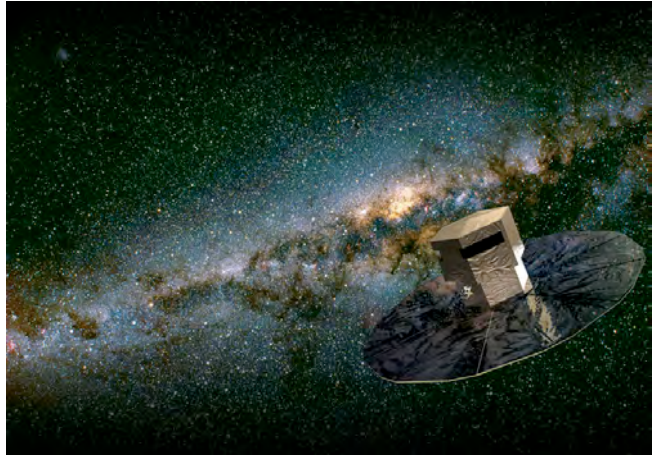

Figure 27. Artist view of the Gaia satellite. From
http://www.rssd.esa.int/index.php?project=GAIA&page=IG_Spacecraft_artistic. Credit: ESA.

### 4.2.3. High Precision Photometry from Space

Space observatories provide particularly favorable observing conditions for high precision photometry: a very stable observing environment allows minimum systematic errors and a precision up to about 100 times better than that of ground-based observations. In addition, observing from space gives the possibility of very long observing runs (up to five months in the case of CoRoT). These are perfect conditions for two fields of research: the detection of transits of exoplanets in front of their parent stars and asteroseismology, i.e. the detection of stellar oscillation modes, providing information about the internal structure of stars.

For a planet to transit in front of its host star, its orbit should be properly aligned with our line of sight. The alignment probability depends of the sizes of the star and of the planet, and of the semi-major axis of the orbit. It is of about 10% for giant planets orbiting their parent star at short distance, decreasing to about 0.5% for an Earth-like planet orbiting a solar-like star at a mean distance of one astronomical unit. It is then essential to be able to observe a very large number of stars to have a chance to detect exoplanet transits. The transit produces a periodic U-shape small drop in stellar light curves, as small as 100 parts per million (ppm) for a terrestrial planet. From the period of the light curve and the duration of the transit, the semi-major axis of the planet's orbit can be determined when the mass of the parent star is preliminarily estimated from its spectral type. The depth of the transit is used to calculate the size of the planet given that the size of the star is preliminary known. Multicolor photometry of the transit allows us to identify false positive detections mimicking planetary transits such as stellar activity or eclipsing binaries. Complementary ground-based radial velocity observations are also made to confirm that the transits are due to planets and to obtain complementary parameters both for the planet (mass, inclination of the orbit compared to the rotational plane of the star) and the parent star (mass, radius and metallicity).

Asteroseismology is the only tool to get information about stellar interiors. Star oscillations are the echo of the internal structure of a star: different oscillation modes penetrate to different depths inside the star. The analysis of the pulsation frequencies (between 1 minute and hundreds of days) and their amplitudes (a few ppm in Fourier space) gives information about the density profile of the regions from where the waves originate and through which they travel: size and composition of the stellar core, limits between the radiative and convective zones, internal profile of rotation.

**CoRoT**

CoRoT (Convection, Rotation and planetary Transits) is a pioneer mission in the domain: it is the first space mission specifically designed for exoplanetary and stellar seismology research. CNES mini-satellite, with contributions from ESA, Austria, Belgium, Germany, Spain and Brazil, it was launched on 27 December 2006 by a Soyuz-Fregat and injected into a low-Earth polar orbit at an altitude of about 900 km. The embarked instrument is a telescope of 27 cm-diameter equipped with a 4-CCD wide-field camera sensitive to tiny variations in the light intensity of stars. The total field of view is about 3 x 2.8 square degrees, one part devoted to asteroseismology targets, the other part to exoplanet transit search. With very long observing runs without interruption (five months) and a photometric accuracy of only a few percent above the photon noise in the seismology channel and twice the photon noise in the exoplanet channel over periods of up to 150 days, CoRoT is measuring the photometric variations of about 200000 stars in two directions of the sky respectively close to the galactic centre and the galactic anticenter. In September 2009, 1000 days after launch, data from over 100000 stars had been acquired and more than 1500 transiting events detected, all over the intervals of period and transit depth accessible to CoRoT. The longest periods, for which a minimum of three transits are detected, are around 75 days. For short or intermediate periods, the large number of transits allows to detect transits close to the theoretical CoRoT detection limit.

As of June 2010, 15 exoplanets have been discovered by CoRoT. The variety of this rich harvest is presented in Figure 28, showing their mass and distance to their parent star. Each newly discovered planet is a new case with different properties: very dense, very large, with a very elongated orbit, etc. Two of them are candidates as being among the very rare inhabitants of the *brown-dwarf desert:* CoRoT-3b and CoRoT-15b. The last one has a mass of about 60 times that of Jupiter and an incredible density of about 40 times that of Jupiter

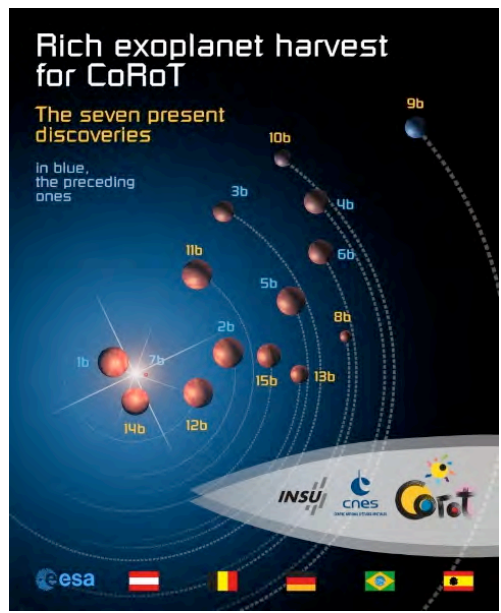

Figure 28. The 15 exoplanets discovered by Corot until June 2010. From http://sci.esa.int/science-e/www/object/index.cfm?fobjectid=47178. Credit: CNES.

The smallest, CoRoT-7b, with a corresponding photometric variation of only 0.03% observed over 153 transits, was the smallest and only rocky exoplanet known in June 2009. It is a Super-Earth with a diameter 1.7 times that of Earth and the shortest revolution period around its host star, about 20 hours. It is probably a member of a new class of planets: super-Earths orbiting extremely close to their parent star. Analysis of very high quality ground-based spectroscopy (mainly obtained by HARPS at the 3.6-mtelescope at La Silla) leads to an estimation of the mass of CoRoT-7b varying between 4.8 and 6.9 Earth-mass, depending on the delicate disentangling of radial velocity variation due to companions (up to three companions are identified) from that due to stellar activity, and on the adopted model atmospheres used to derive the mass and radius of the host star. These results emphasize the importance of complementary ground-based observations and stellar modeling. CoRoT-11b, which discovery was published in June 2010, is an example of the detection capabilities achievable from CoRoT data: the planet is rotating on a prograde orbit with a period of 3 days around a star with a high rotational velocity (rotation period of less than two days). The phase-folded curve of the transit of CoRoT-11b was obtained from the analysis of 49 transits with a depth of about 1 %, occurring every 3 days. The light curves of these two examples are shown in Figure 29.

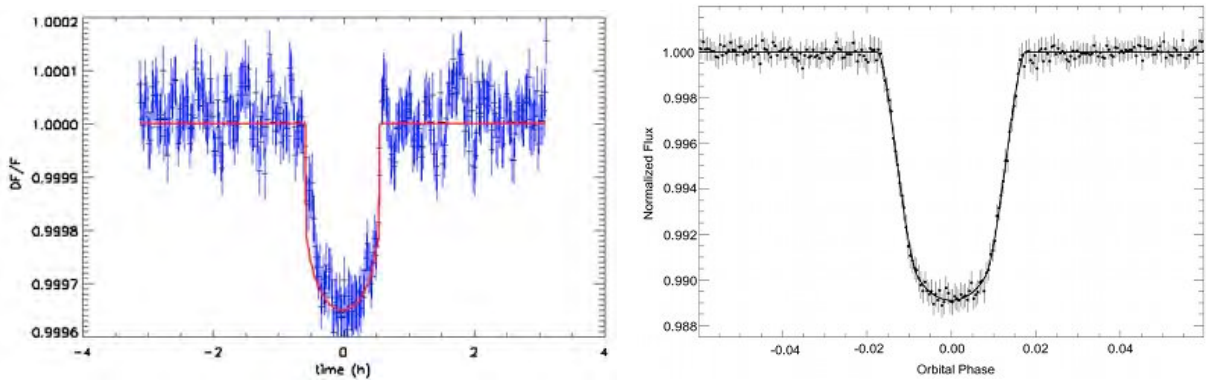

Figure 29. Two examples of folded-light curves obtained by CoRoT.
Left: transit of the smallest exoplanet detected by CoRoT: CoRoT-7b, a small terrestrial planet with a diameter less than twice that of Earth. The parent star is a $V$ = 11.7 G9V star. From http://sci.esa.int/science-e/www/object/index.cfm?fobjectid=44129. Courtesy of A. Léger (A. Léger et al. 2009).
Right: transit of CoRoT-11b: a massive 'hot-Jupiter' in a prograde orbit around a rapidly rotating F6V star of magnitude V=12.9. From http://sci.esa.int/science-e/www/object/index.cfm?fobjectid=47174. Courtesy of D. Gandolfi (Gandolfi et al, 2010).

Already detected in the near infrared by Spitzer Space Telescope, secondary transits, which correspond to the occultation of the planet by its host star, provide precious information about the atmospheric properties of the exoplanet, in particular its albedo. CoRoT's extreme stability and precision allowed the discovery of a secondary transit in the light curve of CoRoT-1b, the first at visible wavelengths. The signal was of order 2/10000. The albedo of CoRoT-1b was found to be small: of the order of 10%, compared to about 40% for the Earth.

A second aspect of the research programme of CoRoT is asteroseismology. Long expected from theory, CoRoT was able to detect for the first time oscillations in solar-like stars, some of them with amplitudes lower than predicted. CoRoT also observed a large variety of other types of stars leading to the determination of their fundamental parameters (mass, radius, luminosity, age, stellar population): red-giant stars, delta Scuti variables, rapidly rotating stars, hot and massive stars, etc.

**Kepler**
Launched in March 2009 by a Delta II rocket from Cape Canaveral, Kepler is a NASA Discovery mission with a 0.95-meter diameter telescope specially designed to monitor 100 to 150000 stars (magnitudes 9 to 16) in a unique field of 115 square degrees in the Cygnus region along the Orion arm, about 12 degrees above the galactic plane and far enough from the ecliptic plane so as not to be obscured by the Sun at any time of the year. Kepler first objective is to detect terrestrial planets in or near the *habitable zone* (region at such a distance from the parent star that liquid water can be maintained on the surface of the planet) for a wide variety of stellar types and to determine their sizes and the characteristics of their orbits. To reach this goal, the telescope was designed to obtain a differential photometric precision of less than 20 ppm (one-sigma) for a 12th magnitude solar-like star observed over 6.5 hours.

From the first 43 days of observations, over one hundred candidate planets, several hundred eclipsing binaries, and thousands of variable stars have been identified in Kepler data. Five new exoplanets with sizes between 0.36 and 1.5 Jupiter radii, masses 0.08 and 2.1 Jupiter mass and orbital periods from 3.2 to 4.9 days were confirmed by radial velocity observations. The temperatures and sizes of the newly discovered planets are shown in Figure 30. The temperatures and sizes of the Solar System planets are indicated for comparison. All five new planets are much hotter than those from the Solar System, and four of them much larger. The five transit light curves are given in Figure 31.

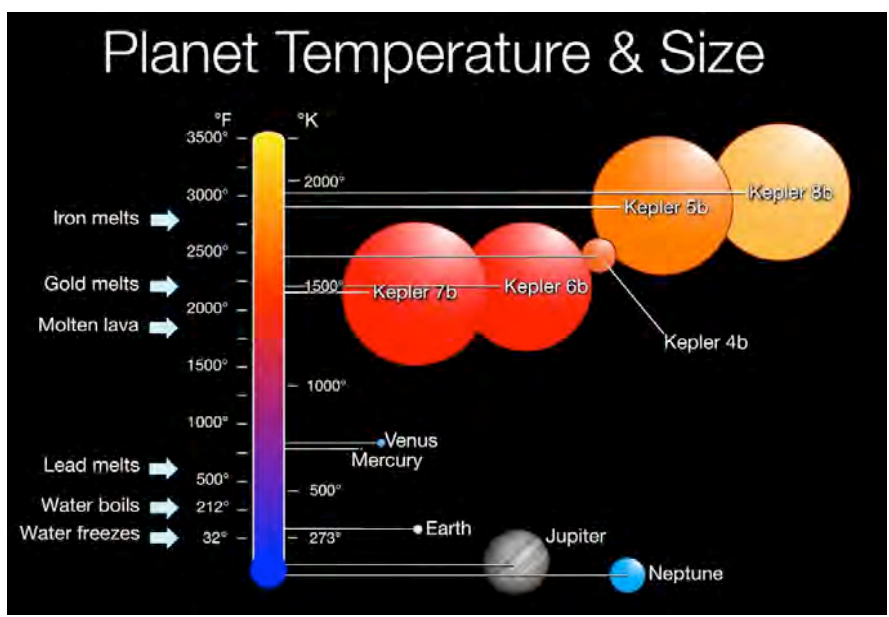

Figure 30. Temperature and size of Kepler first five planets. From
http://kepler.nasa.gov/files/mws/aas2010-3wb-TempAndSize.jpg. Credit: NASA.

Among these first five planets, Kepler-7b has an unusually low density: in January 2010, it was the second lowest density of exoplanets discovered by transits. Its mass is estimated to be smaller than half that of Jupiter (0.43 Jupiter mass) but may even be smaller by as much as 20% because of the possible systematic errors in the preliminary radial velocity measurements. Its radius is about 50% larger than that of Jupiter (1.48 Jupiter radii), leading to a maximum density of 0.17 g cm$^{-3}$. Kepler-7b is orbiting, with a period of about 5 days, a star more massive and much larger than the Sun, which must be near the end of its life on the main sequence.

Kepler-4b is by far the smallest of the series, with the third smallest radius of the planets discovered by transit (0.357 Jupiter radius), it was, in January 2010, the third known transiting Neptune-like planet. It is slightly denser and more massive than Neptune, for about the same size. It is orbiting a G0 star which is near the main sequence turn-off with a period 3.2 days. As shown in Figure 31, the transit is very shallow, with a relative depth of $0.87 \times 10^{-3}$ and a duration of about 3.95 hours. Comparison to models of planets suggests that Kepler-4b has an H/He envelope but is not compact enough to be a water-rich super-Earth.

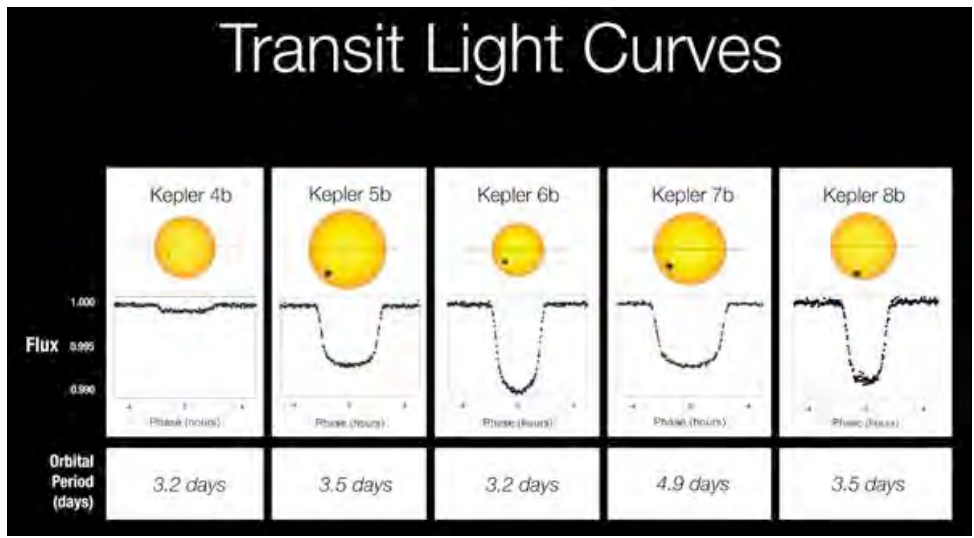

Figure 31. Transit curves of the first five exoplanets discovered by Kepler. From http://kepler.nasa.gov/files/mws/aas2010-1wbLightCurves.jpg. Credit: NASA.

With its high quality photometric observations, Kepler also has a high potential for asteroseismology studies. Two main goals have been established: characterization of the parent stars of discovered planetary systems (radius and age determination) and study of a very large and diverse sample of stars, spanning over main-sequence, subdwarfs, white dwarfs and subgiant stars and a large variety of variable stars and red giants. During the first seven months of mission, a total of 2937 targets have been observed at a 1-min. cadence for periods between 10 days and 7 months. After this initial "Survey Phase" approximately 100 solar-type stars will be selected for long-term observations.

A second-generation mission in this domain, PLATO, is being submitted to ESA as one of the next medium-class missions to be launched in 2017-2018. With a field of view wider than 500 square degrees, it is aiming at providing a full statistical analysis of exoplanetary systems orbiting bright and nearby stars. A photometric precision of 1ppm would allow the detection of Earth-like and super-Earth planets orbiting solar-type stars, the determination of their masses and radius to 1%, and the precise seismic characterization of the host stars.

**Space observatories for the study of dark energy**
A new domain of research for space observatories in the optical and near-infrared wavelengths is being opened for mapping the geometry of the dark Universe and making precise measurements of the expansion rate of the Universe as a function of cosmic time. Two concepts of missions are under study at ESA and NASA, for a possible launch in the 2020 timeframe: Euclid and the Joint Dark Energy Mission (JDEM). With large imaging, photo-z and/or spectroscopic surveys, they are aiming at the measurement of the shapes and redshifts of galaxies and clusters of galaxies, of the light curves of type Ia Supernovae and of baryon acoustic oscillations. All these measurements are designed to lead to a better understanding of the nature of dark energy.

## 5. Infrared and Microwave Space Observatories

The Earth atmosphere is opaque to most of the infrared and microwave wavelengths, absorbed by water vapor, with the exception of a few narrow bands of infrared that can be observed by ground-based observatories. Moreover, our atmosphere, the telescopes and the instruments themselves strongly radiate in the infrared. This is why observations in this wavelength range are made either from a satellite or, for a few wavelengths bands, from telescopes built in high mountains. It is also why detectors, instruments, or even the primary mirror, are cooled to extremely low temperatures, just slightly above absolute zero, in order to eliminate the non-astronomical photon noise and enable the detection and precise measurement of the infrared and microwave photons. The other problem, much further from Earth, is that clouds of gas and dust, very common in the Universe, hide the stars or galaxies embedded in them or situated behind them when they are observed in visible light. Indeed, the wavelength of visible light is comparable to the size of the particles in these clouds and is therefore scattered or absorbed. Infrared and microwave radiation are less affected by these clouds: the longer the wavelength, the thicker the dust cloud that it can penetrate, the cooler the material it can scrutinize.

### 5.1. Infrared Space Observatories

Many objects in the Universe, too cool to be detected in visible light, emit most of their radiation in the infrared wavelengths: comets and asteroids, giant or dwarf cool stars, brown dwarfs and planets, discs of particles around stars, nebulae, dust and gas molecules in the interstellar medium. Regions of formation of galaxies, stars and planets, hidden by dust in the optical wavelengths are also among the first targets of infrared observations. With increased sensitivity, very distant red-shifted galaxies are within reach of the most recent infrared space telescopes, allowing us to probe the early Universe and the formation of the first stars and galaxies. Figure 32 is an illustration of the radiation emitted by a cool star embedded in a dusty grain cloud: the brightness curve has two peaks, one at about 1 micron, maximum emission of the stellar photosphere, the other one at 10 microns, maximum emission of the dust grains surrounding the star.

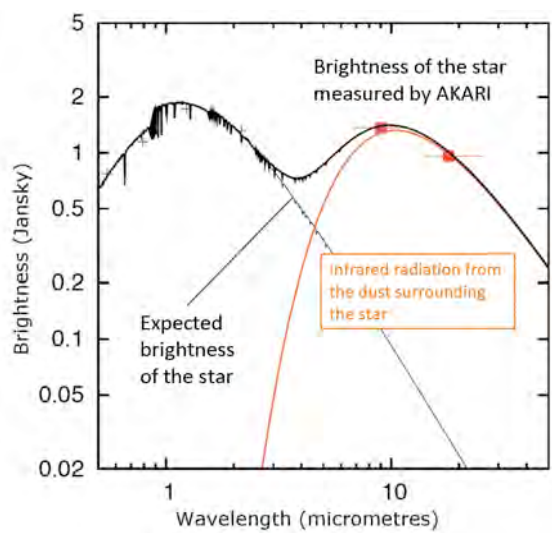

Figure 32. Brightness of a star embedded in dusty material as a function of wavelength measured by AKARI. The cool star has a maximum emission at about 1 micron (thin black curve), the dust grains surrounding the star radiate strongly at wavelengths from 6 to about 20 micrometers (red curve), giving rise to a prominent *infrared excess*. From http://sci.esa.int/science-e/www/object/index.cfm?fobjectid=46755. Credit: JAXA.

Since the early 1980's a series of space infrared observatories (IRAS, ISO, Spitzer, AKARI and WISE) have revolutionized our understanding of the formation and evolution of stars and galaxies thanks to a wealth of observations between wavelengths from about 2 to 200 microns. The first of these satellites, IRAS (Infrared Astronomical Satellite), launched in 1983, was a joint project of the United States (NASA), the United Kingdom and the Netherlands. Helium-cooled telescope of 60 cm, it was operated during 11 months in four broadband photometric channels between 8 and 120 microns, and carried out the first all-sky survey at infrared wavelengths, covering 96% of the sky and detecting about 350000 infrared sources. Twelve years later, ESA's Infrared Space Observatory (ISO), launched in 1995, was operated for 2.5 years in a wider wavelength range (2.5 to 240 microns). With a 60-centimeter diameter primary mirror, it was one thousand times more sensitive and had one hundred times better angular resolution (at 12 μm) compared to IRAS. It discovered signatures of water in many environments, from the atmosphere of planets or in the bright comet Hale-Bopp to star-forming regions, in the vicinity of stars at the end of their lives or in sources very close to the Galactic centre. The three other infrared space telescopes are in orbit in August 2010 and they will be described in little more detail.

**Spitzer**

The Spitzer Space Telescope (formerly SIRTF, the Space Infrared Telescope Facility) is a NASA *Great Observatory* launched from Cape Canaveral on 25 August 2003 in an Earth-trailing orbit. It is a 85-cm telescope designed to operate at infrared wavelengths, from 3 to 180 microns, and equipped with three instruments: the infrared array camera (IRAC) providing simultaneous 5.2 × 5.2 arcmin images at 3.6, 4.5, 5.8, and 8 microns with a spatial resolution of about 2 arcseconds at the shortest wavelengths; the infrared spectrograph (IRS) providing low (R~60-130) and moderate (R~600) spectroscopic resolution from 5.2 to 38 microns and imaging in two filters (13-18 and 18-26 microns); the Multiband Imaging Photometer for Spitzer (MIPS) producing images and photometry in three broad spectral bands, centered at 24, 70, and 160 microns, and low-resolution spectroscopy between 55 and 95 microns. Since May 2009, Spitzer is operating in a "warm" mode with the two shortest wavelength channels of IRAC at 3.6 and 4.5 microns. Spitzer benefited from major technological developments in infrared detectors and lightweight optics. Moreover, its Earth-trailing orbit allows a passive cooling of the telescope to the temperature of 5.5 K, the liquid helium cryostat being only used to cool the instruments to temperatures of 1.4 K.

A wealth of results has been obtained from the satellite data on many topics involving cool objects and dust-obscured regions, ultra-luminous infrared galaxies and very remote objects red-shifted to the infrared. A few examples are presented below. Figure 33 is a three-color composite image of the inner part of our Galaxy, covering about 50% of the galactic plane. It was obtained by two of Spitzer instruments: IRAC observations in two wavelengths (3.6 microns in blue, 8 microns in green) and MIPS (24 microns, in red). This survey results from more than 800000 individual frames and shows the various cool components of the galactic plane: the light from stars is in blue, polycyclic aromatic hydrocarbons organic molecules in green, warm dust in red, star-forming regions in red and yellow (showing the presence of both dust and polycyclic aromatic hydrocarbons). The Orion nebula, a very active area of star formation, including very young stars and blown away surrounding dust, is shown in Figure 34. Figure 35 was obtained by using all three instruments on-board Spitzer (3.6 microns in blue, 8 microns in green, 24 microns, in red). It shows the distribution of stars and polycyclic aromatic hydrocarbons all over the M101 galaxy. The outer red zones lack organic molecules present in the rest of the galaxy. These dusty molecules are usually present in star formation zones and are believed to contribute to the cooling of the clouds from which stars are formed. The outer part of M101 is therefore a good place to investigate how star formation can happen without these molecules, as was the case in the early Universe.

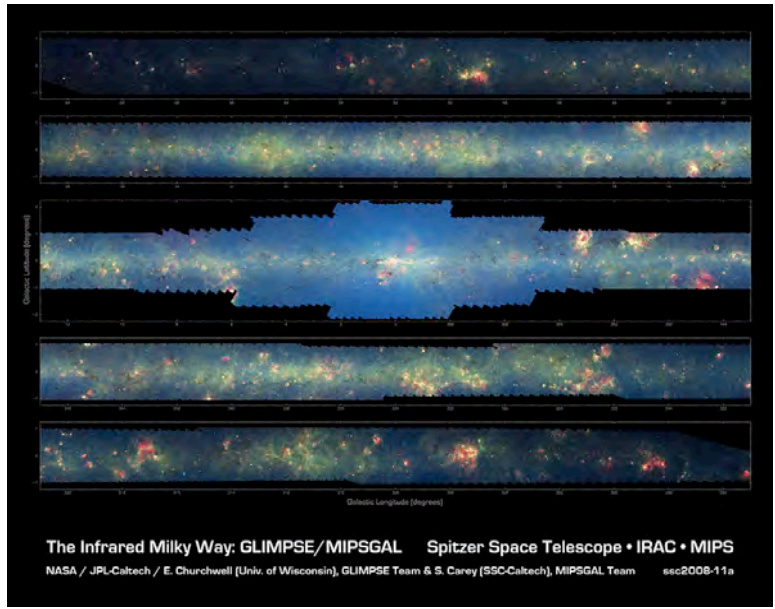

Figure 33. Spitzer mosaic of the inner Milky Way. In this mosaic the galactic plane is broken up into five components: the far-left side of the plane (top image); the area just left of the galactic centre (second to top); galactic centre (middle); the area to the right of galactic centre (second to bottom); and the far-right side of the plane (bottom). From http://www.spitzer.caltech.edu/images/2680-ssc2008-11a-Spitzer-Finds-Clarity-in-the-Inner-Milky-Way. Credit: NASA/JPL-Caltech/Univ. of Wisconsin.

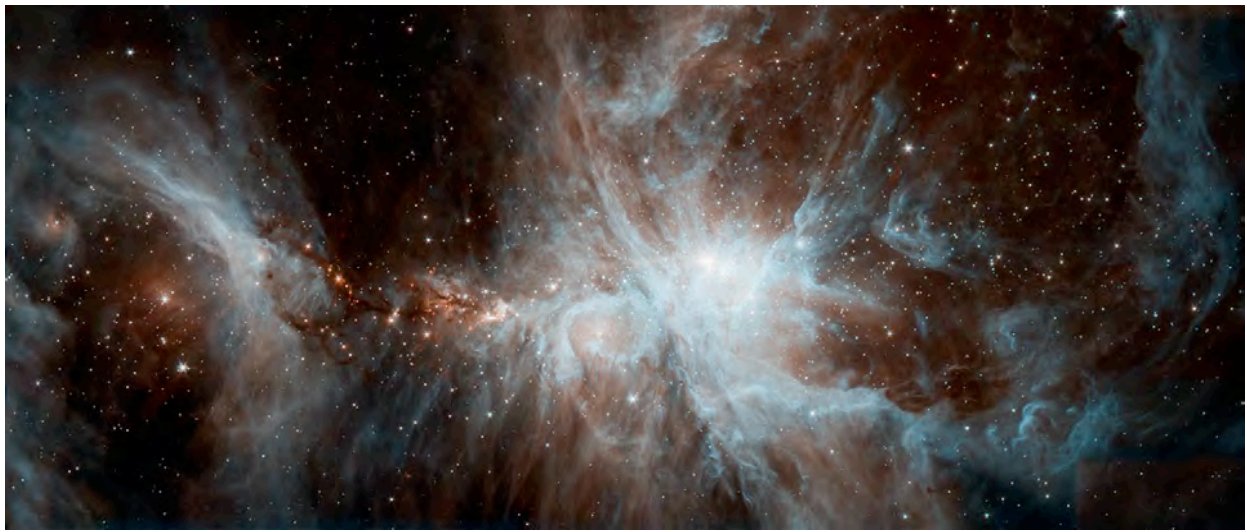

Figure 34. Spitzer image of the Orion nebula, taken by the IRAC camera during the warm part of the mission (April 2010). Blue: 3.6-micron image. Orange: 4.5-micron light image. From http://www.spitzer.caltech.edu/images/3018-sig10-003-Orion-s-Dreamy-Stars. Credit: NASA/JPL-Caltech/J. Stauffer.

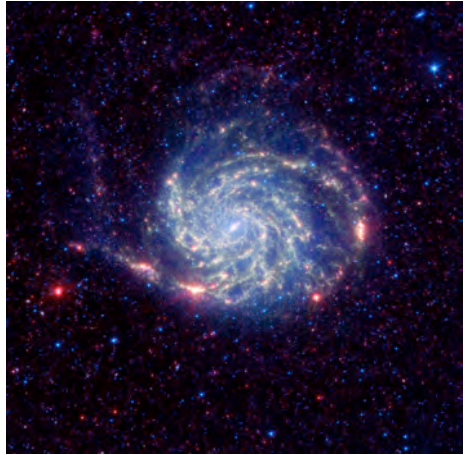

Figure 35. Spitzer image of the M101 galaxy in the constellation Ursa Major, about 11 Mpcs away. From http://www.spitzer.caltech.edu/images/1943-ssc2008-14a-No-Organics-Zone-Circles-Pinwheel-Galaxy. Credit: NASA/JPL-Caltech/K. Gordon, STScI.

**AKARI**

AKARI, launched in February 2006 in a sun-synchronous orbit at an altitude of 700 km, is the first Japanese infrared astronomical satellite. With a 68.5-cm telescope cooled down to 6 K, it was observing in the 1.7 to 180 microns wavelength range and surveyed the entire sky in six infrared bands between May 2006 and August 2007. The resulting catalogues contain the positions and flux values for more than 1.3 million celestial sources detected by the two instruments carried by AKARI: the Infrared Camera (IRC) detected about 870 000 objects in the mid-infrared (9 and 18 microns), and the Far-Infrared Surveyor (FIS) detected about 430 000 celestial sources in four far-infrared bands (65, 90, 140, and 160 microns). AKARI also performed more than 5000 pointed observations over the wavelength range 2-180 microns in 13 bands, providing multi-wavelength observations of nearby solar system objects, young stars and debris discs, brown dwarfs, evolved stars in our Galaxy, and a dedicated survey of the Large Magellanic Cloud. The AKARI all-sky map is shown in Figure 36. Stars dominate at 9 microns (blue) whereas dust and star formation regions are highlighted at 90 microns (red). At the end of August 2007, AKARI run out of liquid helium and started its post-helium phase (operating at about 40 K) in which imaging and spectroscopic capabilities are only available in the 1.8 to 5.5 micron wavelength range.

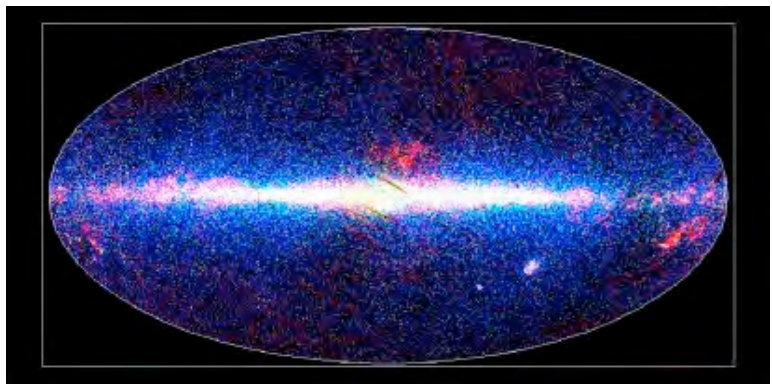

Figure 36. AKARI all-sky map at 9 microns (blue), 18 microns (green) and 90 microns (red). The plane of the Galaxy is the prominent feature, running horizontally across the map, with the Galactic centre in the middle. From http://sci.esa.int/science-e/www/object/index.cfm?fobjectid=46756. Credit: JAXA.

**WISE**

WISE (Wide-field Infrared Survey Explorer) is a NASA Explorer mission launched in December 2009 in circular 525-km Sun-synchronous orbit. With a 40-cm wide angle telescope cooled to 15K and equipped with new Megapixel infrared detector arrays (resolution between 6 arcsec at the shortest wavelengths to 12 arcsec at 22 microns), it achieved a shallow survey of the entire sky in the mid-infrared in four bands (3.4, 4.6, 12 and 22 microns) in July 2010. Early August 2010, the satellite ran out of hydrogen and started to warm up, however still observing in the shortest infrared band, less sensitive to heat. The survey includes 1.5 million observations and discovered millions of objects, including asteroids, stars and galaxies. The most interesting of these objects will be targets for follow-up studies with other missions also observing in the infrared domain such as Spitzer, Herschel or JWST. A mosaic image of the Pleiades cluster obtained by combining data from the four WISE detectors is shown in Figure 37. Stars are easily distinguished from warm dust by the multi- wavelengths observations.

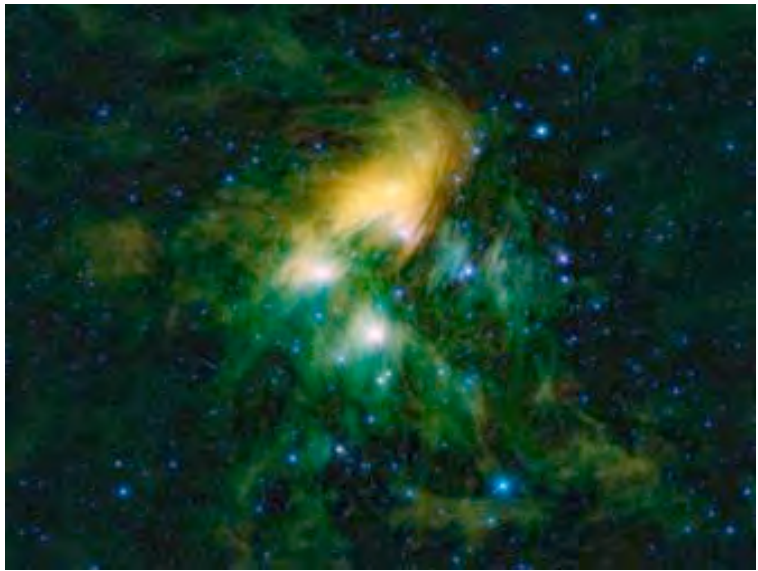

Figure 37. Mosaic image of the Pleiades obtained by the four infrared detectors of WISE in February 2010. Blue and cyan: infrared light at 3.4 and 4.6 microns, dominated by stars. Green and red: light at 12 and 22 microns, mostly from warm dust. From http://photojournal.jpl.nasa.gov/catalog/PIA13121. Credit: Credit: NASA/JPL-Caltech/UCLA.

**JWST**

The James Webb Space Telescope (JWST), known initially as the Next Generation Space Telescope (NGST), is a collaborative project between NASA, ESA and CSA (the Canadian Space Agency), planned for launch by an Ariane 5 rocket into a Sun-Earth L2 halo orbit no sooner than 2014. It will be the largest telescope ever sent in space with a passively cooled 6.5-m lightweight deployable primary mirror and four science instruments primarily designed to work in the near-infrared, from 2 to 5 microns, with extensions into the visible (0.6 to 2 microns) and mid- infrared (5 to 28 microns). Compared to the Hubble Space Telescope (2.4-m mirror diameter and observing from the near-infrared to the far-ultraviolet), the JWST observatory is clearly optimized for studies of the faint extragalactic Universe back in time and redshift to the epoch of the ignition of the very first stars. Nonetheless, it will also be a general-purpose observatory, but no servicing missions will be possible in the chosen orbit, particularly stable and favorable for an astronomy spacecraft, but situated 1.5 million kilometers away from Earth. An artist view of the observatory is shown in Figure 38.

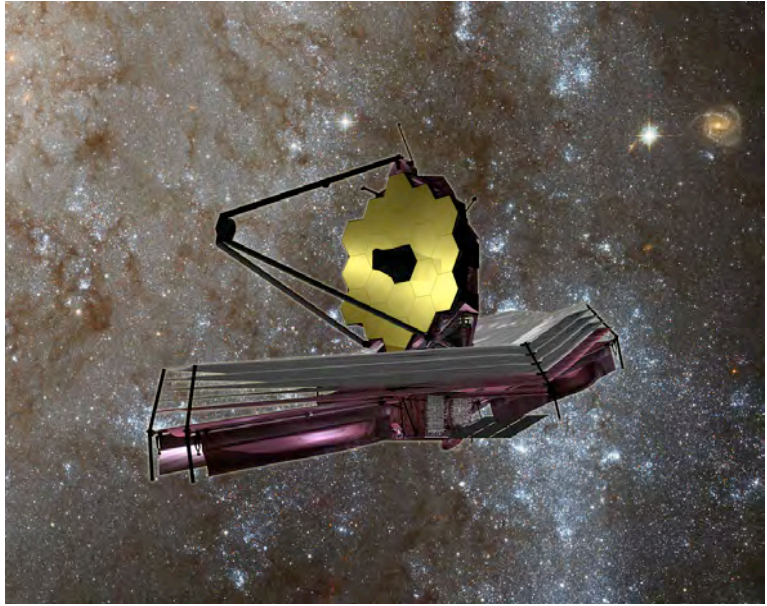

Figure 38. An artist view of the James Webb Space Telescope. From
http://www.stsci.edu/jwst/overview/gallery.html. Credit: NASA- STScI.

**JWST will carry four science instruments:**
- NIRCam (Near-Infrared Camera), a near-IR and visible camera covering wavelengths from 0.6 to 5 microns with a wide field of view (2.2 x 4.4 arcmin) and coronagraphic capabilities;
- NIRSpec (Near-Infrared Spectrograph), a wide field (3.4 x 3.4 arcmin) multi-object near-IR spectrometer capable to observe more than 100 objects simultaneously and covering wavelengths from 0.6 to 5 microns at spectral resolutions of about 100, 1000 and 2700;
- MIRI (Mid-Infrared Instrument), a combined mid-IR camera (1.3 x 1.9 arcmin) and long-slit spectrograph (spectral resolution about 3000) covering wavelengths from 5 to 28 microns, with coronagraphic capabilities;
- FGS (Fine Guidance System), a system which will allow a milli-arcsecond level pointing stability and a precise guiding at any point on the sky (2.2 x 2.2 arcmin field of view, operating between 1.5 and 5 microns.

Several innovative technologies have been developed for JWST. The first challenge is the 6.5-m mirror: made of 18 monolithic ultra-lightweight beryllium individual mirrors with a thin gold coating, it will be folded so that it can fit into the rocket, and will be unfolded and adjusted to the desired shape after launch. Each of the individual hexagonal mirror segments is 1.3 m in diameter and weights approximately 20 kg. Six of these modules are shown in Figure 39. Many other new technologies are being developed and tested such as the very sensitive and low-noise large-area infrared detector arrays, the microshutters being used for NIRSpec that open and close when a magnetic field is applied allowing to view or block a portion of the sky, or the cryocooler providing the possibility to operate the mid-IR detectors at the temperature of 7K.

JWST is designed to examine every phase of cosmic history, from the first light after the Big Bang to the formation of galaxies, stars, and planets to the evolution of our own solar system. The science goals are presented in four themes: the end of the dark ages, identification of the first bright objects that formed in the early Universe and the history of reionization; the assembly of galaxies (including our own) over the cosmic time and how they evolved to their present status; the structure, formation and evolution of stars and the formation of protoplanetary discs and of planetary systems; the characterization of the surface, atmospheric, and surrounding discs composition of planets and exoplanets and the search for origins of life.

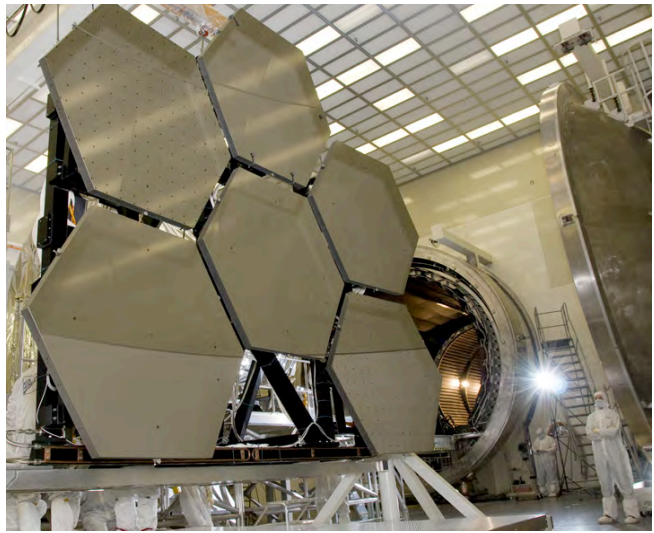

Figure 39. Six of the 18 James Webb Space Telescope mirror segments are being prepared for a cryogenic test at NASA's Marshall Space Flight Centre. Five days are needed to cool a mirror segment to -25K, permitting engineers to measure in extreme detail how the shape of each mirror changes as it cools. From http://www.jwst.nasa.gov/images_mirror30.html. Credit: NASA/MSFC/Emmett Givens.

**Future infrared space observatory projects**

Several projects of infrared space observatories are under study in the various space agencies over the world. In the near infrared and optical wavelengths, two concepts of missions have already been mentioned in section 4.2, aiming at the elucidation of one of the most challenging recent discoveries: dark energy. A new concept of wide-field infrared survey telescope (WFIRST) has been proposed by the US Decadal Survey by mid-August 2010. Its ambition is to jointly tackle two different domains: on the one hand, the acceleration of the expansion rate of the Universe and dark energy by combining a week lensing survey, the detection of baryon acoustic oscillations and of supernova explosions; and on the other hand, the frequency of various types of exoplanetary systems, especially Earth-like planets, by performing a microlensing census. At longer wavelengths, SPICA (SPace Infrared telescope for Cosmology and Astrophysics) is a JAXA led mission (with considered contributions from ESA and NASA) in the mid- to far-infrared wavelengths (5 to 210 microns) building on the success of AKARI, Spitzer and Herschel. The concept includes a cooled 3.5-meter aperture telescope like Herschel, but actively cooled to below 5 K to eliminate all non-astronomical photon noise and allow a breakthrough in detection sensitivity. Proposed for a launch in the 2018 timeframe, it is expected to provide imaging and spectroscopic capabilities for probing galaxy, star and planetary formation and the evolution of dust and gas in the interstellar medium of galaxies (from our own to very distant galaxies). A coronagraph is also proposed to obtain mid infrared uncontaminated spectra of young massive planets.

**5.2. Far-Infrared - Sub-Millimeter Space Observatories**

ESA's Herschel Space Observatory (formerly called FIRST, Far Infrared and Sub-millimeter Telescope) is the first space observatory to study the Universe radiating in the far infrared to sub-millimeter radiation domain (55-672 microns), bridging a gap between earlier infrared space missions and ground-based facilities. With a 3.5-m primary mirror, the largest single mirror ever built for use in space, Herschel is opening a new window, almost unexplored, towards some of the coldest and most distant objects in the Universe: objects with temperatures between 5 and 50 K which have radiative emission peaks in this radiation range, and gases with temperatures between ten and a few hundred K which have their brightest molecular and atomic emission lines at these wavelengths. Fourth *cornerstone* mission in the ESA Science Programme, it was launched on 14 May 2009, jointly with Planck, by an Ariane 5 rocket and directly injected into a Lissajous orbit around the L2 point (second Lagrange point of the Sun-Earth system) at a distance of around 1.5 million km from Earth. With technically challenging telescope and instruments, the Herschel Space Observatory was developed by ESA, with collaborations with scientific institutes in ESA member states, Canada and the USA.

Herschel carries a telescope passively cooled to 80 K with a 3.5-m diameter silicon carbide (SiC) mirror. The mirror was in itself a challenge: huge but very light (320 kg), made out of 12 segments which have been brazed together to form a monolithic mirror of about 3 mm thickness, later polished to obtain a surface accuracy of 1 micron and recovered of a reflective aluminum layer accounting for only about 10 g (!) of the total weight. The telescope is equipped with three complementary instruments: two cameras/medium resolution spectrometers (PACS - Photodetector Array Camera and Spectrometer and SPIRE - Spectral and Photometric Imaging Receiver) and a very high-resolution heterodyne spectrometer (HIFI - Heterodyne Instrument for the Far Infrared), all three housed in a superfluid helium cryostat to keep them at temperatures between 2 and 15 K. The detectors of PACS and SPIRE have even to be cooled down to 0.3 K. The choice of an orbit around L2 provides a stable thermal environment and a total absence of atmospheric emission or absorption. Combined with a large mirror, very low telescope emissivity and very performing instruments, Herschel offers very sensitive large-scale imaging photometric capability in six bands with centre wavelengths from 70 to 500 microns, medium resolution spectroscopy with limited imaging capability over the entire Herschel wavelength coverage, and high to very high resolution spectroscopy over much of the coverage.

Designed to observe the cool universe from our Solar System to newly forming stars in our Galaxy and through to star-forming galaxies out to very high redshifts, Herschel already provided, after only a few months of observation, spectacular new results presented in a special issue of the review Astronomy & Astrophysics (July 2010). Two domains can be highlighted: in our Galaxy, where detailed chemical analysis of the interstellar medium is providing a wealth of new information about the life cycle of matter and where stars forming in very dusty regions can now be detected; and in very high redshift galaxies where very deep surveys reveal star forming regions at cosmological distances. A few examples are given below.

A very detailed, high-resolution, spectrum of a very prolific star formation region in our Galaxy, the Orion Nebula, has been obtained by the HIFI instrument. It is shown in Figure 40, superimposed on a Spitzer image of Orion. The spectral richness is impressive, showing, among many organic molecules, water, carbon monoxide, formaldehyde or methanol. These data are the promise of a much deeper understanding of the chemistry associated with star and planetary formation and will allow us to derive very detailed models of the star-forming clouds.

Figure 41 presents a composite image of the Rosette molecular cloud where previously unseen very massive stars (bright white points), up to ten times the mass of the Sun, are embedded inside opaque dusty cocoons. Lower mass protostars, similar in mass to the Sun, are also detected (redder regions). This is a good illustration of the observing capacity of the PACS and SPIRE instruments, making images in different far-infrared wavelengths.

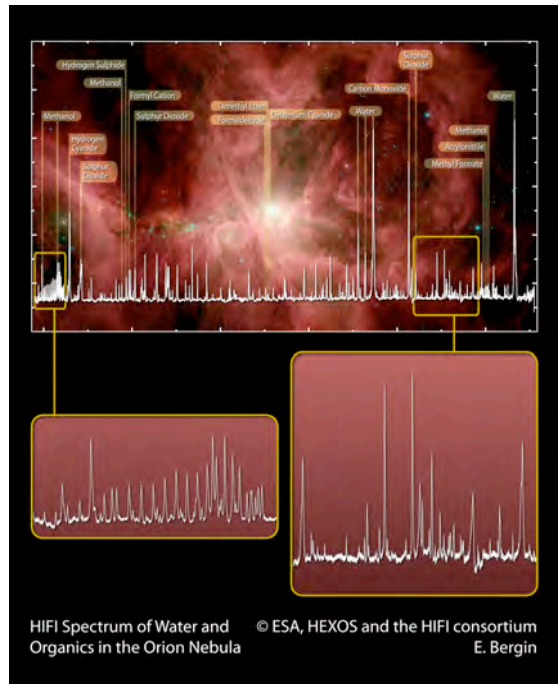

Figure 40. Herschel-HIFI spectrum of the Orion Nebula, superimposed on a Spitzer image of Orion. The spectrum illustrates the spectral richness of regions of star and planet formation. From http://sci.esa.int/science-e/www/object/index.cfm?fobjectid=46653. Credit: ESA, HEXOS and the HIFI Consortium.

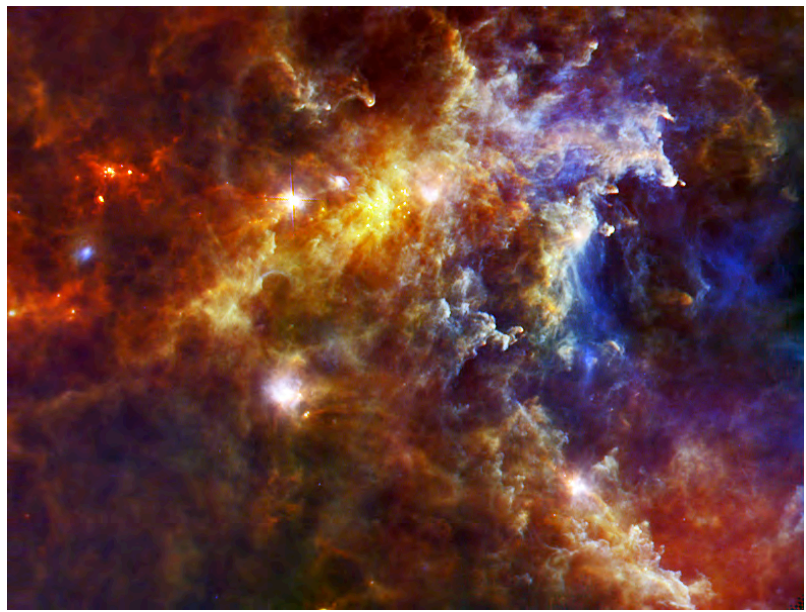

Figure 41. Herschel-PACS and SPIRE composite image of the Rosette molecular cloud, taken at three different infrared wavelengths: 70 microns (blue), 160 microns (green) and 250 microns (red). The different colors also correspond to different temperatures of dust, from 10K for the red emission, to 40K in the blue. From http://sci.esa.int/science-e/www/object/index.cfm?fobjectid=46822. Credit: ESA/PACS & SPIRE Consortium/HOBYS Key Programme.

The Herschel Space Telescope also pointed its instruments to the *GOODS* (Great Observatories Origins Deep Survey) fields, two carefully selected regions of the northern and southern sky, centered on the Hubble Deep Field North and the Chandra Deep Field South, and already observed over a broad wavelength range by many ESA's and NASA's space observatories and ground-based telescopes. These two fields of 10 x 16 arcmin have been specially chosen for the study of galaxies out to very high redshifts, as they are free from any strong emission from the Milky Way (stars or interstellar matter). These new observations in the far-infrared domain, unexplored at these sensitivity, reveal a previously unresolved population of galaxies responsible for the diffuse Cosmic Infrared Background (CIB), isotropic emission with a peak around 100-200 microns associated with the formation of galaxies. A composite image of a subset of the GOODS-North field, obtained in the three SPIRE bands is presented in Figure 42. Every fuzzy blob in this image is a very distant galaxy, seen as it was 3-10 thousand million years ago. The properties of these galaxies appear surprisingly uniform over this time span, suggesting a simple and universal mechanism for star formation.

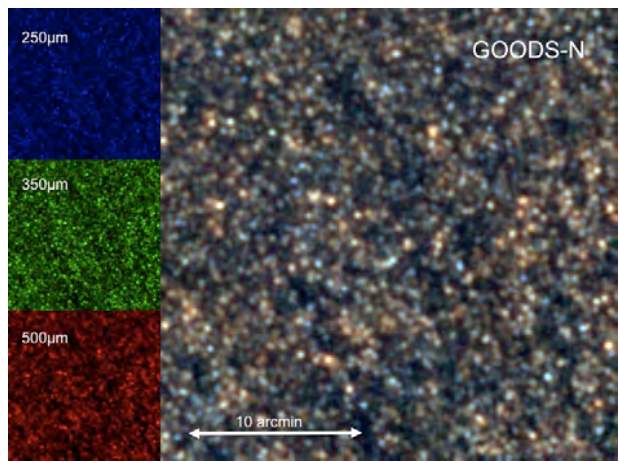

Figure 42. Herschel-SPIRE composite image of the GOODS-North field in the three SPIRE bands (red: 500 microns, green: 350 microns, blue: 250 microns): an area 30 x 30 arc minutes showing a myriad of very distant galaxies. From http://sci.esa.int/science-e/www/object/index.cfm?fobjectid=46974. Credit: ESA/SPIRE Consortium/HerMES Key Programme Consortium.

## 5.3 Millimeter-Sub-Millimeter Space Observatories

Astronomy in the millimeter and sub-millimeter domain only began in the 1960's, opening a very rich field of research: the study of the cold and very cold Universe. Millimeter-wave bands are the best suited for tracing molecular material, even in very tenuous regions, both nearby and very far away, back in time to the early Universe (when it was only 380000 years old). Last but not least, the Cosmic Microwave Background (CMB), fossil radiation released soon after the Big-Bang, has a spectrum peaking in the mm-wave band at the present epoch. Theoretically predicted in the 1950's, it was discovered by chance by two radio astronomers, Penzias and Wilson, who were developing very sensitive microwave receivers. They received the Nobel Prize in 1978 for their discovery. The CMB fills the entire sky and initially appeared to be very uniform. Only a careful analysis of the radiation, observed with high angular resolution and sensitivity, has revealed tiny variations, seeds from which galaxies and clusters of galaxies grew from the hydrogen/helium gas that was spread across the entire Universe at the time the radiation was released.

Ground-based observation of these high frequency radio waves (ranging from about 0.3 to 10 mm) is limited by Earth's atmosphere emission (increasing the level of noise) and absorption by water vapor (restricting observations to a few wavelength windows). This is why millimeter and sub-millimeter radio-telescopes are usually situated in high mountains (for example ALMA, the Atacama Large Millimeter/sub-millimeter Array, observing from 0.3 mm to 3.6 mm, is located in the Chilean Andes, 5000 m above sea level). High-altitude balloons and aircrafts are also used.

However, only space observatories can provide the necessary homogeneity, sensitivity and spatial resolution to measure the tiny fluctuations of the CMB, identify regions very slightly warmer or colder than the average and precisely map their distribution all over the sky. COBE (Cosmic Background Explorer) was the first of these Cosmic Microwave Background space observatories. Developed by NASA to measure the diffuse infrared and microwave radiation, it was launched in November 1989 and carried three instruments: DIRBE (Diffuse Infrared Background Experiment) was designed to map the infrared sky brightness in the wavelength range 1.25 to 240 microns, DMR (Differential Microwave Radiometer) made differential measurements between two directions in the sky separated by 60 degrees in three wavelengths (9.5, 5.7 and 3.3 mm) and over the entire sky, and FIRAS (Far Infrared Absolute Spectrometer) precisely measured the cosmic microwave background spectrum and observed the dust and line emission from the Galaxy. COBE discovered the intrinsic anisotropy of the CMB, at a level of a part in 100000, and showed that its spectrum is that of a nearly perfect blackbody with a temperature of 2.725 ± 0.002 K.

**WMAP**
WMAP (Wilkinson Microwave Anisotropy Probe) is the second generation of CMB space observatories. It was launched by NASA in June 2001 into a Lissajous orbit about the L2 Sun-Earth Lagrange point, 1.5 million km from Earth, providing a very stable thermal environment and near 100% observing efficiency since the Sun, Earth, and Moon are always behind the instrument's field of view. WMAP is a scanning satellite rapidly covering the sky with two, back to back, symmetric telescopes, measuring the temperature difference between two areas on the sky roughly 140° apart, and producing full sky maps of the CMB temperature anisotropy. The instrument has five frequency bands (13.6, 10.0, 7.5, 5.0 and 3.3 mm) to facilitate separation of galactic foreground signals from the cosmic background radiation. Large primary reflectors (1.4 x 1.6 meter diameter), sensitive receivers and severe control of systematic errors provide the desired angular resolution and sensitivity (45 times the sensitivity and 33 times the angular resolution of the COBE): an angular resolution of at least 0.3° over the full sky, a sensitivity of 20 μK per 0.3° square pixel, with systematic errors limited to 5 μK per pixel. The overall goal is measuring the temperature of the microwave sky to an accuracy of one millionth of a degree. WMAP ceased science operations on 19 August 2010. Results from the final two years of data will be published in approximately two years.

The results from seven years of observations were released in January 2010, leading to the fine-resolution (0.2°) full-sky map of the CMB sky shown in Figure 43. The signal from our Galaxy was subtracted using the multi-frequency data. This image shows temperature fluctuations in a temperature range of ± 200 μK. From these results, the age of the Universe was estimated to be 13.73 billion years to within 1% (0.12 billion years), the curvature of space determined to be within 1% of the flat Euclidian model, and the respective proportions of baryons (4.6 ± 1.3 %), dark matter (23.3 ± 1.3 %) and dark energy (72.1 ±1.5 %) re-evaluated more precisely. Polarization anisotropies are also much better determined in this new data release.

Figure 44 provides the 7-year temperature power spectrum from WMAP (dots and error bars) compared with the ΛCDM model (Lambda-Cold Dark Matter model = *standard* model) best fit to the 7-year WMAP data (red line). The *angular power spectrum* of the anisotropy of the CMB is a plot of how much the temperature varies from point to point on the sky (the *y*-axis variable: relative brightness of the small areas) as a function of the angular frequency (the *x*-axis variable *l*: the larger is *l* the smaller is the area with similar temperature). The agreement between the observed quantities and the *standard* model is remarkable down to the third peak of the curve. The shape of this curve contains a wealth of cosmological information: curvature of the Universe, baryon and dark mater density, etc.

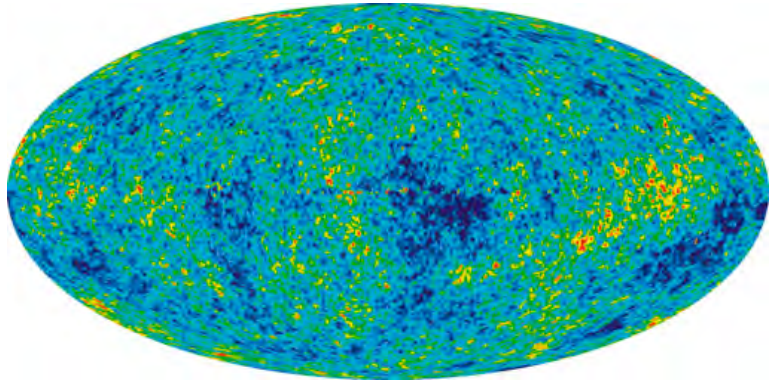

Figure 43. All-sky map created from seven years of WMAP data showing temperature fluctuations: warmer areas are red, cooler areas are blue. From http://map.gsfc.nasa.gov/media/101080/index.html. Credit: NASA / WMAP Science Team.

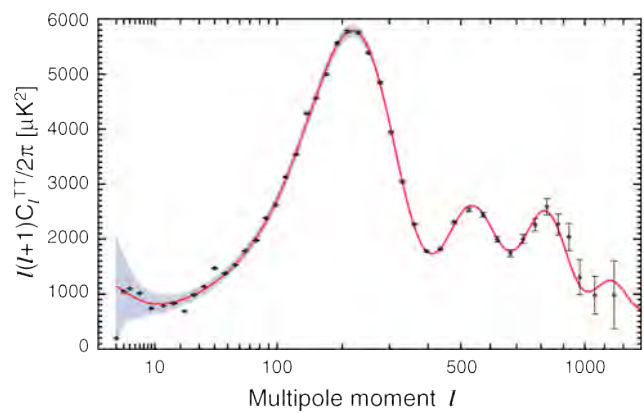

Figure 44. The 7-year temperature power spectrum from WMAP (black dots and error bars), compared to the ΛCDM model best fit (red curve) to the data. From http://lambda.gsfc.nasa.gov/product/map/current/pub_papers/sevenyear/powspectra/wmap_7yr_power_spectra_images.cfm. Credit: NASA / WMAP Science Team.

**Planck**
Planck, named after the German Nobel laureate Max Planck (1858-1947), is the third generation of CMB space observatories and the first European space observatory in this domain. Included in the ESA's Scientific Programme in 1996, it was launched on 14 May 2009, jointly with Herschel by an Ariane 5 rocket. The two spacecraft separated soon after launch and now operate independently in two different Lissajous orbits around the L2 point. With a 1.9 x 1.5 m telescope, Planck focuses microwave radiation from the sky onto two powerful cryogenic instruments, operating at extremely low temperature: LFI (the Low Frequency Instrument), an array of radio receivers based on HEMT (High Electron Mobility Transistor Mixers) tuned amplifiers, covering the frequency range 27-77 GHz (11–3.9 mm), and operated at a temperature of 20 K; and HFI (the High Frequency Instrument), an array of bolometric detectors, covering the frequency range 100–857 GHz (3–0.35 mm), and operated at a temperature of 0.1 K). Combining the two instruments, the sky is mapped over nine bands of wavelengths, from microwaves to the far-infrared (one centimeter to 350 microns), enabling the separation of the Galactic and extragalactic foreground radiations (dust and gas from the interstellar medium, free-free and synchrotron emissions) from the primordial cosmological background signal (microwave and far-IR). Thanks to the above characteristics, the angular resolution of Planck maps is greater than 5 arcminutes and the temperature resolution of the order of one part in $10^6$.

Before the start of routine operations, a *First Light Survey* was performed in August 2009, covering a two-week period during which Planck surveyed the sky continuously. It yielded maps of rings about 15 degrees wide, stretching across the full sky, one for each of Planck's nine frequencies. It demonstrated the stability of the instruments and the ability to calibrate them over long periods to the exquisite accuracy needed. From the end of this First Light Survey, by the end of August 2009, Planck is continuously scanning the sky, making a full all-sky map in approximately 6 months. It is expected that Planck will be able to perform four complete all-sky scans before the end of cryogenic operations.

In July 2010, the first all-sky image of Planck was published, highlighting the two major sources of microwave radiations: the cosmic background and the diffuse emission from our Galaxy. Figure 45 is a composite image, in galactic coordinates, synthesized from data covering the whole Plank frequency range. The tiny temperature fluctuations of the CMB, reflecting the primordial density variations, are clearly visible at high galactic latitudes where the foreground contribution from the Galaxy is not predominant. At lower galactic latitudes but also well away from the galactic plane, the map is dominated by the diffuse emission originating from gas and dust in the Milky Way, showing the extent of our Galaxy's large-scale structure and its emission properties. Some particularly intense sources are noted in the map: in the Galaxy (the Galactic Centre, the giant molecular clouds of Perseus and Orion which are extremely active regions of star formation), in the Local Group (the two Magellanic Clouds and Andromeda), and radio sources at extragalactic distances (Centaurus A and 3C 273). Superimposed on the Galactic Plane are the contours of the densest molecular clouds traced by a survey of carbon monoxide (CO), identifying regions of intense star formation. The rich texture of the cold dust, showing loops and spurs, and the differences in temperature within the galactic interstellar medium are illustrated in Figure 46 by an image combining two Planck channels (540 and 350 microns) and data from IRAS at 100 microns. The temperature differences are reflecting different physical status of the interstellar medium, tracing the evolution of star formation.

Although the various components of the microwave sky appear to be separable only in certain areas of the sky, sophisticated image analysis techniques, made possible by Planck frequency coverage and by the unprecedented resolution and accuracy of its measurements, will allow us to isolate each of these components and reveal in great details the underlying map of the CMB.

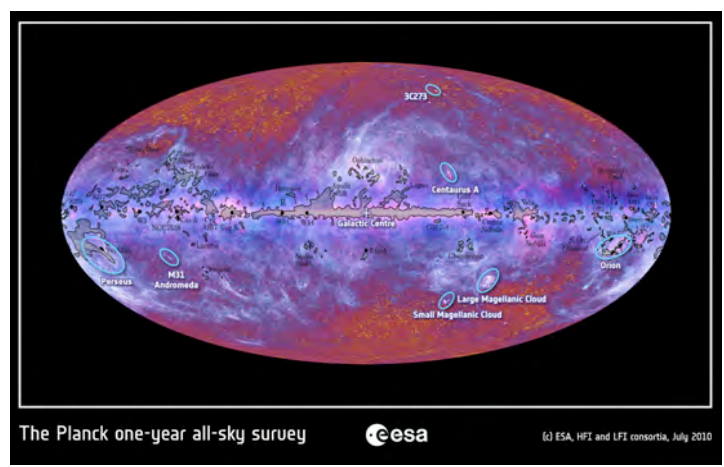

Figure 45. Composite multi-color all-sky image of the microwave sky obtained from the full wavelength range of Planck from 10 mm to 350 microns: it is dominated by the CMB at high galactic latitudes and by diffuse emission from the Galaxy at low and intermediate latitudes. From http://sci.esa.int/science-e/www/object/index.cfm?fobjectid=47341. Credit: ESA, HFI and LFI consortia; CO map from T. Dame et al., 2001.

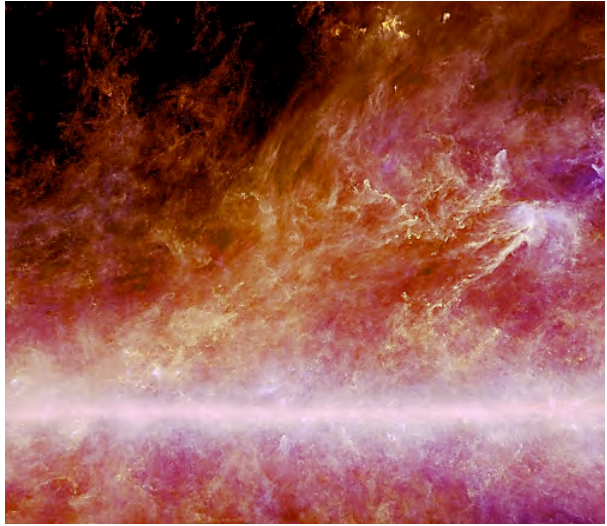

Figure 46. Image combining two Planck channels (540 and 350 microns) and data from IRAS at 100 microns. The image shows the large-scale structure of the coldest dust particles in the Milky Way where massive stars are forming. Red corresponds to temperatures as cold as 12 K, white to warmer areas (a few tens of degrees). From http://sci.esa.int/science-e/www/object/index.cfm?fobjectid=46707. Credit: ESA and the HFI Consortium, IRAS.

## 6. Gravitational Waves Space Observatories

Another completely new window to extreme phenomena may be open by LISA, a fundamental physics project jointly developed by ESA and NASA, in the gravitational waves frequency range from $10^{-4}$ to $10^{-1}$ Hz, for a possible launch in 2020. The two main categories of celestial objects that should be sources of gravitational waves in this domain are galactic binaries and massive black holes expected to exist in the centers of galaxies. The most powerful sources of such waves are the mergers of massive black holes with masses of $10^6$ to $10^8$ solar masses in the centers in distant galaxies. Besides testing the theory of General Relativity, the detection of gravitational waves would provide information about the unknown phenomena taking place in the surroundings of massive black holes, and about their formation and growth.

## 7. Conclusion

This chapter is a very selective presentation of the wealth of results obtained from observatories in space. Each of them opens a new window to the Universe, exploring the high variety of celestial objects from different perspectives: emission in new wavelength ranges all over the electromagnetic spectrum, sharper images, much higher photometric or astrometric accuracies, new possibilities of full-sky surveys or extremely long non-interrupted observations. A few characteristics have been highlighted in all sections: the power of discovery of homogeneous observations of very large parts of the sky and of all-sky surveys, possibly leading to the unexpected at the occasion of serendipitous observations; the complementarities of multi-wavelength, multi-scale observations obtained both from space and ground-based observations; the importance of deep/ultradeep fields, pushing the performances of an observatory to its extreme limits.

New space missions are being studied which will allow observing further back in time, closer to extreme phenomena, deeper into the dust, discovering still new types of exoplanets and analyze their properties in great detail. With larger telescopes, more sensible receivers, higher time resolution, larger fields of view, the new instruments will provide answers to present questions … and open many others. A few such projects or concepts have been quoted in each of the above sections. Each of them is the result of scientific motivations open by previous observatories, in space or on ground, and of the development of advanced technologies: lightweight mirrors, filters and optical elements; lightweight grazing angle reflectors for high energy telescopes; miniaturized components; smaller detections systems with better radiation tolerance and wider dynamic range; detectors operating in thermal vacuum, cryogenic or passively cooled environments; more sensitive polarimeters; better shielding to radiations; etc. These technology developments very often have further applications in the everyday life, in aeronautics, informatics or medicine for example.

Besides obtaining fantastic science results, working in such projects is an inspiring and demanding adventure, asking for rigour and pragmatism, enthusiasm and realism, and a great school to learn or develop how to work with colleagues with different cultures, languages, or background.

**Acknowledgements**
It is a pleasure for me to acknowledge Didier Barret and Frédéric Arenou for their very careful reading of the manuscript, and Bozena Czerny, the editor of this chapter of the UNESCO Encyclopedia, for her sharp comments.

Intensive use has been made of the SAO/NASA Astrophysics Data System (ADS).

**Glossary**

**CMB:** cosmic microwave background, electromagnetic radiation of the temperature of 2.7 K, filling the Universe. It is the relic of the early epoch when the Universe was hot and fully ionized.

**OB associations:** groups of young massive stars.

**Halo:** roughly spherical component of a galaxy, extending beyond the main visible part of the galaxy.

**HR diagram:** Hertzsprung–Russell diagram. A graph showing the relation between absolute magnitude, or luminosity of stars, and their temperatures, colors or spectral types.

**Astroseismology:** studies of the stellar internal structure through observation and modeling the stellar pulsations.

**Dark energy:** hypothetical form of energy responsible for the rapid expansion of the Universe

**Binary systems:** gravitationally bound two celestial bodies (usually stars)

**Quasar:** bright and distant active galactic nucleus.

**AGN:** active galactic nuclei. In those nuclei other processes than stellar lead to emission of radiation. Their activity is explained as being powered by accretion onto massive black holes.

**Accretion disc:** a structure formed by gas in orbital motion around a central body.

**Supernovae:** very energetic stellar explosions at the late stage of stellar evolution. Several types of supernovae are recognized, considered to be driven by different internal mechanism.

**Seyfert galaxies:** - one of the types of active galactic nuclei.

**Baryon acoustic oscillations:** overdensity of baryonic matter at certain length scales due to acoustic waves propagating in the early Universe.

**Photon noise:** electronic noise appearing in the device due to the statistical fluctuations in the number of discrete charges carrying the current.

**Planetary nebula:** - emission nebula arising from the expanding shell of ionized gas ejected by a star, which is at the asymptotic giant branch evolutionary stage.

**Intensive use was made of the web sites of**

ESA:
- http://sci.esa.int
- http://sci.esa.int/science-e/www/area/index.cfm?fareaid=1#,
- http://www.rssd.esa.int/index.php?project=ASTRONOMY,
- http://xmm.esac.esa.int
- http://herschel.esac.esa.int/

NASA:
- http://www.nasa.gov/mission_pages
- http://heasarc.gsfc.nasa.gov/docs/observatories.html
- http://fermi.gsfc.nasa.gov/
- http://kepler.nasa.gov/
- http://www.jwst.nasa.gov
- http://map.gsfc.nasa.gov/

Astronomy and astrophysics
- Special issue: *"Herschel: the first science highlights"*, Vol. 518, July-August 2010

Cambridge X-ray Astronomy
- http://www-xray.ast.cam.ac.uk

CNES
- http://smsc.cnes.fr/COROT/index.htm

Hubble
- http://www.spacetelescope.org

Jet Propulsion Laboratory
- http://www.jpl.nasa.gov/
- http://www.spitzer.caltech.edu/

Smithsonian Center for Astrophysics
- http://chandra.harvard.edu

Space Telescope Science Institute
- http://www.stsci.edu/
- http://www.stsci.edu/jwst/
- http://archive.stsci.edu/missions.html

**Reference of this review paper:**

Catherine Turon, 2010, Observatories in Space, review 50 pages and 46 figures, in Astronomy and Astrophysics, Eds. Oddbjorn Engvold, Rolf Stabell, Bozena Czerny, John Lattanzio, in Encyclopedia of Life Support Systems (EOLSS), Developed under the Auspices of the UNESCO, Eolss Publishers, Oxford, UK, [http://www.eolss.net].